\documentclass[a4paper,12pt]{article}
\pdfoutput=1
\usepackage{amssymb}
\usepackage{amsmath}
\usepackage{amsfonts}
\usepackage{mathtools}
\usepackage{bm}
\usepackage{slashed}
\usepackage{color}
\usepackage{indentfirst}
\usepackage{bbm}
\usepackage{csquotes} 
\usepackage{caption}
\usepackage{cite}
\usepackage{here}
\usepackage{comment}
\usepackage{listings}
\usepackage[italicdiff]{physics}
\numberwithin{equation}{section} 

\usepackage[unicode]{hyperref}
\usepackage{xcolor}
\hypersetup{
    colorlinks=false,
    citebordercolor=lime,
    linkbordercolor=pink,
}

\def\ignore#1{{}}

\makeatletter
\newcommand*\rel@kern[1]{\kern#1\dimexpr\macc@kerna}
\newcommand*\widebar[1]{%
  \begingroup
  \def\mathaccent##1##2{%
    \rel@kern{0.8}%
    \overline{\rel@kern{-0.8}\macc@nucleus\rel@kern{0.2}}%
    \rel@kern{-0.2}%
  }%
  \macc@depth\@ne
  \let\math@bgroup\@empty \let\math@egroup\macc@set@skewchar
  \mathsurround\z@ \frozen@everymath{\mathgroup\macc@group\relax}%
  \macc@set@skewchar\relax
  \let\mathaccentV\macc@nested@a
  \macc@nested@a\relax111{#1}%
  \endgroup
}
\makeatother

\newcommand{\brkt}[1]{\left( #1 \right)}
\newcommand{\brc}[1]{\left\{ #1 \right\}}
\newcommand{\sbk}[1]{\left[ #1 \right]}
\newcommand{\tl}[1]{\tilde{#1}}

\renewcommand{\thefootnote}{\fnsymbol{footnote}}

\begin{document}

\title{
\begin{flushright}
\begin{minipage}{0.25\linewidth}
\normalsize
KEK-TH-2490 \\
KYUSHU-HET-253 \\*[50pt]
\end{minipage}
\end{flushright}
{\Large \bf 
Full higher-dimensional analysis of moduli oscillation and radiation in expanding universe
\\*[20pt]}}

\author{
Hajime~Otsuka$^{a}$\footnote{
E-mail address: otsuka.hajime@phys.kyushu-u.ac.jp
}
\ and\
Yutaka~Sakamura$^{b,c}$\footnote{
E-mail address: sakamura@post.kek.jp
}\\*[20pt]
$^a${\it \normalsize
Department of Physics, Kyushu University,}\\ 
{\it \normalsize
744 Motooka, Nishi-ku, Fukuoka, 
819-0395, Japan}\\
$^b${\it \normalsize 
KEK Theory Center, Institute of Particle and Nuclear Studies, KEK,}\\
{\it \normalsize 1-1 Oho, Tsukuba, Ibaraki 305-0801, Japan}\\
$^c${\it \normalsize 
Graduate University for Advanced Studies (Sokendai),}\\
{\it \normalsize 1-1 Oho, Tsukuba, Ibaraki 305-0801, Japan.}
}
\date{}

\maketitle

\centerline{\small \bf Abstract}
\begin{minipage}{0.9\linewidth}
\medskip 
\medskip 
\small

We investigate effects of the radiation and the moduli oscillation around the stabilized values on the evolution 
of a 6-dimensional spacetime compactified on $S^2$. 
In order to see the transition from the 5-dimensional space to the 3-dimensional one, 
we develop a procedure to pursue the spacetime evolution with appropriate approximations, 
which is valid until the spacetime behaves like 4-dimensional. 
In the case that the moduli stabilization process cannot be described in the context of the 4-dimensional effective theory, 
it takes quite a long time for the moduli oscillation to dominate the total energy density, 
in contrast to the conventional result obtained by the 4-dimensional effective theory approach. 
We also found that even if the moduli are set at the stabilized values, 
they start to oscillate due to the pressure in the extra space~$S^2$ in some cases. 
\end{minipage}

\renewcommand{\thefootnote}{\arabic{footnote}}
\setcounter{footnote}{0}
\thispagestyle{empty}
\clearpage
\addtocounter{page}{-1}


\section{Introduction}
Moduli fields naturally appear in the low-energy effective theory of higher-dimensional gravity. 
Since couplings of matter fields and the spacetime evolution will depend on a dynamics of moduli fields, 
they will give impacts on several aspects of particle physics and cosmology.

In the conventional approach, the dynamics of moduli fields will be treated 
in the framework of four-dimensional (4D) effective field theories (EFT) under the assumption that Kaluza-Klein (KK) modes
 are decoupled from low-energy dynamics.
However, such an assumption will be violated when the typical moduli mass~$m$ is the same order of or larger than 
the mass of the first KK excitation mode~$m_{\rm KK}^{(1)}$, i.e., $m \gtrsim m_{\rm KK}^{(1)}$. 
This situation is realized in the large volume regime of the extra-dimensional space. 
Although such a situation is not disfavored for any specific reasons, most of the works about the moduli dynamics have been 
studied in the context of 4D EFT.

As a simple setup to analyze a higher-dimensional gravity theory, we consider the so-called Salam-Sezgin model \cite{Salam:1984cj}, 
which is based on a gauged six-dimensional (6D) supergravity~\cite{Nishino:1984gk,Randjbar-Daemi:1985tdc} 
compactified on a sphere with a U(1) magnetic flux. 
The original work for this model discussed a static background. 
Its time-dependent extension was discussed in Refs. \cite{Maeda:1984gq} in the radiation-dominated universe, which is 
described by the hypermultiplets and/or the vector multiplets~\footnote{Note that the cancellation of 6D gravitational anomalies 
requires the existence of hypermultiplets and/or vector multiplets \cite{Green:1984bx}. 
}. 
Since there is a flat direction in the moduli space in the original Salam-Sezgin model, we introduce a dilaton potential 
so that all the moduli are completely fixed. 
Then, the moduli oscillate around their stabilized values in the evolution of the universe. 
In the 4D EFT approach, it is well-known that the energy density of the moduli oscillation rapidly dominates over that of the radiation, 
and the expanding space behaves like the 4D matter-dominated universe~\cite{Coughlan:1983ci}-\cite{deCarlos:1993wie}. 
However, this is not the case when the moduli stabilization procedure cannot be described in the 4D EFT. 
In our previous work~\cite{Otsuka:2022rpx}, we investigated the evolution of the background spacetime during the
moduli stabilization process that is assumed to occur in the radiation dominated era.
It was numerically found that when the modulus mass is larger than the KK mass, 
the radiation contribution to the total energy density remains non-negligible for a long time in contrast to the conventional 4D EFT analysis.
However, this numerical analysis is available for only a limited range of the time and the parameters, 
and it will be difficult to see the transition from 6D to 4D explicitly.
The purpose of this paper is to analytically investigate effects of the radiation and the moduli oscillation on the spacetime evolution,  
especially focusing on a parameter region in which one cannot use the 4D EFT analysis.
Our findings are summarized as follows:
\begin{itemize}
    \item The radiation remains non-negligible for a long time when the moduli stabilization procedure cannot be described 
    in the framework of 4D EFT, namely $m \gtrsim m_{\rm KK}^{(1)}$. 
   
    \item In such a case, for lower initial temperatures, the universe never experiences the moduli-dominated era 
    if the moduli decay before the dominance of the moduli oscillation.
   
    \item Even if the moduli are set at the stabilized values, they start to oscillate due to the pressure in the extra space $S^2$ 
    in some parameter spaces.
\end{itemize}
For our purpose, we develop a procedure to compute various quantities at late times 
that enables us to pursue the transition from 6D to 4D. 

The paper is organized as follows. 
In the next section, we provide a brief review of the model used in our previous work~\cite{Otsuka:2022rpx}. 
In Sec.~\ref{Approximations}, we explain how to compute various quantities at later times. 
In Sec.~\ref{sec:conditions}, we discuss the conditions that the radiation dominates the total energy density. 
Sec.~\ref{sec:con} is devoted to the summary. 
In the appendices, we provide brief derivations of some formulae used in the text, 
and show the conservation law of the energy-momentum tensor.

\section{Setup}
\label{sec:2}
In this section, we briefly review the model we considered in our previous work~\cite{Otsuka:2022rpx}. 
The whole spacetime is 6D, and a 2D subspace is compactified on a sphere~$S^2$. 
The indices~$M,N=0,1,2,\cdots,5$ denote the 6D coordinate ones, $\mu,\nu=0,1,2,3$ denote the 4D ones 
for the non-compact space, and $m,n=4,5$ are the 2D ones for the compact space. 
As the coordinates on $S^2$, we choose the spherical ones~$(x^4,x^5)=(\theta,\phi)$, 
where $\theta$ and $\phi$ are the polar and the azimuthal angles, respectively.

\subsection{Model inspired by 6D supergravity on a sphere}
The model is given by~\footnote{
We work in the unit of the 6D Planck mass. 
}
\begin{align}
 S &= \int d^6x\;\sqrt{-g^{(6)}}\brc{-\frac{1}{2}R^{(6)}-\frac{1}{2}\partial^M\sigma\partial_M\sigma-\frac{g^2e^{\sigma}}{4}F^{MN}F_{MN} -V(\sigma)}, 
\end{align}
where $R^{(6)}$ denotes the 6D Ricci scalar, $\sigma$ is a real scalar (dilaton), 
$F_{MN}\equiv\partial_MA_N-\partial_NA_M$ is the field strength of the U(1) gauge field~$A_M$, 
and $g$ is the gauge coupling constant. 
The scalar potential~$V(\sigma)$ is given by 
\begin{align}
 V(\sigma) &= 2e^{-\sigma}+\frac{m^2}{2}(\sigma-\sigma_*)^2, 
 \label{def:Vscalar}
\end{align}
where $m$ and $\sigma_*$ are positive constants. 

This is basically the bosonic part of the gauged 6D ${\cal N}=(1,0)$ supergravity~\cite{Nishino:1984gk,Randjbar-Daemi:1985tdc}, 
except for the following two points. 
First, we drop the self-dual antisymmetric tensor field~$B_{MN}$ because it is irrelevant to the following discussions. 
Second, we add the second term in (\ref{def:Vscalar}) to the scalar potential in order to stabilize the moduli completely. 

The equations of motion are
\begin{align}
 R_{MN}^{(6)}-\frac{1}{2}g_{MN}R^{(6)}-T_{MN}^{\rm matter} &= 0, \nonumber\\
 \frac{1}{\sqrt{-g^{(6)}}}\partial_M\brkt{\sqrt{-g^{(6)}}\partial^M\sigma}
 -\frac{g^2e^\sigma}{4}F^{MN}F_{MN}- \partial_\sigma V(\sigma) &= 0, \nonumber\\
 \partial_M\brkt{\sqrt{-g^{(6)}}e^\sigma F^{MN}} &= 0, 
 \label{EOM}
\end{align}
where the energy-momentum tensor~$T_{MN}^{\rm matter}$ is 
\begin{align}
 T_{MN}^{\rm matter} &= -\partial_M\sigma\partial_N\sigma+\frac{1}{2}g_{MN}\partial^L\sigma\partial_L\sigma
 -g^2e^\sigma F_{ML}F_N^{\;\;L}+\frac{g^2e^\sigma}{4}g_{MN}F^{PQ}F_{PQ}+g_{MN}V(\sigma). 
\end{align}

In the absence of the second term in (\ref{def:Vscalar}), this model has the following static background~\cite{Salam:1984cj}: 
\begin{align}
 g_{\mu\nu} &= \eta_{\mu\nu} = {\rm diag}\,(-1,1,1,1), \nonumber\\
 g_{44} &= b^2, \;\;\;\;\;
 g_{45} = g_{54} = 0, \;\;\;\;\;
 g_{55} = b^2\sin^2\theta, \nonumber\\
 F_{\mu\nu} &= F_{\mu m} = 0, \nonumber\\
 F_{45} &= -F_{54} = \frac{\sin\theta}{2g}, \;\;\;\;\;
 F_{44} = F_{55} = 0, \nonumber\\
 \sigma &= \ln(4b^2),  \label{bg:SS}
\end{align}
where $\eta_{\mu\nu}$ is the 4D Minkowski metric, and $b$ is a positive constant. 
In this case, the constant~$b$ is a free parameter, 
and the size of the compact space~$S^2$ remains to be unfixed. 

In the presence of the second term in (\ref{def:Vscalar}), the background~(\ref{bg:SS}) remains to be a solution, 
but now $\sigma$ is fixed to the constant~$\sigma_*$. 
Hence the constant~$b$ is also fixed as
\begin{align}
 b &= b_* \equiv \frac{e^{\sigma_*/2}}{2}.  \label{bvac}
\end{align}

In addition to the above field content, we also introduce the radiation contribution. 
In the context of the 6D ${\cal N}=1$ supergravity, 
the number of hypermultiplets~$n_H$ and that of vector multiplets~$n_V$ are constrained 
by the anomaly cancellation condition~$n_H-n_V=244$~\cite{Randjbar-Daemi:1985tdc,Green:1984bx,Kumar:2010ru}.\footnote{
The number of tensor multiplet is assumed to be one, otherwise the theory cannot be described by the Lagrangian. 
} 
This indicates that a large number of hypermultiplets must exist in a consistent theory. 
We assume that scalars in such hypermultiplets do not have nontrivial background values, 
but they contribute to the radiation that fills in the whole 5D space. 
Then, the energy-momentum tensor appearing in the Einstein equation must include the radiation contribution: 
\begin{align}
 \brkt{T^{\rm rad}}_M^{\;\;N} &= \begin{pmatrix} \rho^{\rm rad} & & & \\
 & -p^{\rm rad}_3{\bm 1_3} & & \\ & & -p^{\rm rad}_2 & \\ & & & -p^{\rm rad}_2\sin^2\theta \end{pmatrix}, 
\end{align}
where $\rho^{\rm rad}$, $p^{\rm rad}_3$ and $p^{\rm rad}_2$ are the radiation energy density, 
the pressures in the non-compact 3D space and in the compact 2D space, respectively, 
whose explicit forms are listed in Appendix~\ref{TDquantities}. 
In the presence of the radiation, the static background~(\ref{bg:SS}) and (\ref{bvac}) are no longer a solution of the equations of motion, 
and the universe continues to expand. 
Thus we make the following ansatz for the background. 
\begin{align}
 g_{MN} &= \begin{pmatrix} -1 & & & \\
 & a^2(t){\bm 1_3} & & \\ & & b^2(t) & \\ & & & b^2(t)\sin^2\theta \end{pmatrix}, \nonumber\\
 F_{\mu\nu} &= F_{\mu m} = 0, \nonumber\\
 F_{45} &= -F_{54} = \frac{\sin\theta}{2g}, \;\;\;\;\;
 F_{44} = F_{55} = 0, \nonumber\\
 \sigma &= \sigma(t), 
 \label{bg:ansatz}
\end{align}
where $a(t)$ and $b(t)$ are the scale factors for the non-compact 3D space and the compact 2D space, respectively.

\subsection{Evolution equations}
Under the background ansatz~(\ref{bg:ansatz}), the equations of motion in (\ref{EOM}) become 
\begin{align}
 \frac{3\dot{a}^2}{a^2}+\frac{\dot{b}^2}{b^2}+\frac{6\dot{a}\dot{b}}{ab}+\frac{1}{b^2}
 -\frac{1}{2}\dot{\sigma}^2-\frac{e^\sigma}{8b^4}-V(\sigma)-\rho^{\rm rad} &= 0, \nonumber\\
 \frac{2\ddot{a}}{a}+\frac{\dot{a}^2}{a^2}+\frac{2\ddot{b}}{b}+\frac{\dot{b}^2}{b^2}+\frac{4\dot{a}\dot{b}}{ab}
 +\frac{1}{b^2}+\frac{1}{2}\dot{\sigma}^2-\frac{e^\sigma}{8b^4}-V(\sigma)+p^{\rm rad}_3 &= 0, \nonumber\\
 \frac{3\ddot{a}}{a}+\frac{3\dot{a}^2}{a^2}+\frac{\ddot{b}}{b}+\frac{3\dot{a}\dot{b}}{ab}+\frac{1}{2}\dot{\sigma}^2
 +\frac{e^\sigma}{8b^4}-V(\sigma)+p^{\rm rad}_2 &= 0, \nonumber\\
 \ddot{\sigma}+\brkt{\frac{3\dot{a}}{a}+\frac{2\dot{b}}{b}}\dot{\sigma}+\frac{e^\sigma}{8b^4}+\partial_\sigma V(\sigma) &= 0, 
 \label{bg:EOM}
\end{align}
where the dot denotes the time derivative. 
The first equation is the $(t,t)$-component of the Einstein equation. 
Since this does not contain the second order $t$-derivatives, it is regarded as a constraint on the initial conditions 
of the time evolution. 
The second and the third equations come from the diagonal components for the 3D non-compact space and the 2D compact space, 
respectively. 
The other components of the Einstein equation vanish. 
The last equation is the dilaton field equation. 
In addition to these, we can obtain the evolution equation for the inverse temperature~$\beta$ 
from the conservation law, as shown in Appendix~\ref{conserv_law}.  

In the following, we assume that the chemical potential~$\mu$ is negligible, i.e., $\beta\mu\ll 1$. 
Then, if we redefine the scale factors as
\begin{align}
 A \equiv \ln a, \;\;\;\;\;
 B \equiv \ln b, 
\end{align}
the above equations are rewritten as
\begin{align}
 \ddot{A} =& -\frac{9}{4}\dot{A}^2+\frac{1}{4}\dot{B}^2-\frac{1}{2}\dot{A}\dot{B}-\frac{1}{8}\dot{\sigma}^2
 +\frac{1}{2}\brkt{e^{-\frac{\sigma}{2}}-\frac{e^{\frac{\sigma}{2}-2B}}{4}}\brkt{e^{-\frac{\sigma}{2}}+\frac{3e^{\frac{\sigma}{2}-2B}}{4}}
 \nonumber\\
 &+\frac{m^2}{8}\brkt{\sigma-\sigma_*}^2+\frac{p^{\rm rad}_3-2p^{\rm rad}_2}{4}, \nonumber\\
 \ddot{B} =&\; \frac{3}{4}\dot{A}^2-\frac{7}{4}\dot{B}^2-\frac{3}{2}\dot{A}\dot{B}-\frac{1}{8}\dot{\sigma}^2
 +\frac{1}{2}\brkt{e^{-\frac{\sigma}{2}}-\frac{e^{\frac{\sigma}{2}-2B}}{4}}\brkt{e^{-\frac{\sigma}{2}}-\frac{5e^{\frac{\sigma}{2}-2B}}{4}} 
 \nonumber\\
 &+\frac{m^2}{8}\brkt{\sigma-\sigma_*}^2-\frac{3p^{\rm rad}_3-2p^{\rm rad}_2}{4}, \nonumber\\
 \ddot{\sigma} =&-\brkt{3\dot{A}+2\dot{B}}\dot{\sigma}
 +2\brkt{e^{-\frac{\sigma}{2}}-\frac{e^{\frac{\sigma}{2}-2B}}{4}}\brkt{e^{-\frac{\sigma}{2}}+\frac{e^{\frac{\sigma}{2}-2B}}{4}}
 -m^2\brkt{\sigma-\sigma_*}, 
 \label{evolv_eq1}
\end{align}
with the constraint:~\footnote{
This $\hat{\rho}^{\rm tot}$ is related to the 6D total energy density~$\rho^{\rm tot}$ defined in (\ref{comp:T}) as 
$\hat{\rho}^{\rm tot}=\rho^{\rm tot}-\frac{1}{2}\dot{\sigma}^2-e^{-2B}$, where the last term corresponds to the curvature of $S^2$. 
} 
\begin{align}
 3\dot{A}^2+\dot{B}^2+6\dot{A}\dot{B}-\frac{1}{2}\dot{\sigma}^2 &= 2\brkt{e^{-\frac{\sigma}{2}}-\frac{e^{\frac{\sigma}{2}-2B}}{4}}^2
 +\frac{m^2}{2}\brkt{\sigma-\sigma_*}^2+\rho^{\rm rad} \nonumber\\
 &\equiv \hat{\rho}^{\rm tot}. 
 \label{cstrt}
\end{align}
Using (\ref{cstrt}) and (\ref{rel:rhop}), the second equation in (\ref{evolv_eq1}) can be rewritten as
\begin{align}
 \ddot{B} &= -\brkt{3\dot{A}+2\dot{B}}\dot{B}+\brkt{e^{-\frac{\sigma}{2}}-\frac{e^{\frac{\sigma}{2}-2B}}{4}}
 \brkt{e^{-\frac{\sigma}{2}}-\frac{3e^{\frac{\sigma}{2}-2B}}{4}}+\frac{m^2}{4}\brkt{\sigma-\sigma_*}^2+p_2^{\rm rad}. 
 \label{evolv_eq:B}
\end{align}

The energy density and the pressures are expressed as (see Appendix~\ref{TDquantities}) 
\begin{align}
 \rho^{\rm rad} &= \frac{g_{\rm dof}e^{-2B}}{8\pi^3\beta^4}\brc{\pm 6{\rm Li}_4(\pm 1)+3Q_1+Q_2}, \nonumber\\
 p^{\rm rad}_3 &= \frac{g_{\rm dof}e^{-2B}}{8\pi^3\beta^4}\brc{\pm 2{\rm Li}_4(\pm 1)+Q_1}, \nonumber\\
 p^{\rm rad}_2 &= \frac{g_{\rm dof}e^{-2B}}{16\pi^3\beta^4}Q_2, 
 \label{expr:rhop}
\end{align}
where $g_{\rm dof}$ is the degrees of freedom for 6D relativistic particles, 
$\beta$ is the inverse temperature, 
the functions~$Q_i(x)$ ($i=1,2,3$) are defined in (\ref{def:Q_1}), (\ref{def:Q_2}) and (\ref{def:Q_3}), 
and their arguments are $\beta/b=\beta e^{-B}$.  
The upper (lower) signs represent the case that the radiation consists of the bosons (fermions). 
The evolution equation for $\beta$ is obtained from (\ref{eq_for_beta}) as 
\begin{align}
 \frac{\dot{\beta}}{\beta} &= \frac{3\dot{A}\brc{\pm 8{\rm Li}_4(\pm 1)+4Q_1+Q_2}+\dot{B}\brkt{2Q_2+Q_3}}
 {\pm 24{\rm Li}_4(\pm 1)+12Q_1+5Q_2+Q_3}. 
 \label{evolv_eq2}
\end{align}

The equations~(\ref{evolv_eq1}) and (\ref{evolv_eq2}) with the constraint~(\ref{cstrt}) are the evolution equations 
for the expanding 5D space. 
Note that
\begin{align}
 {\rm Li}_4(1) &= \zeta(4), \;\;\;\;\;
 {\rm Li}_4(-1) = -\frac{7}{8}\zeta(4). 
\end{align}
In the following, we consider a case that the radiation consists of only the bosonic particles 
to simplify the discussion. 
Hence the upper signs in (\ref{expr:rhop}) and (\ref{evolv_eq2}) are applied.

For numerical computation, we choose the initial conditions at $t=0$ as
\begin{align}
 a(0) &= 1, \;\;\;\;\;
 b(0) = b_{\rm I}, \;\;\;\;\;
 \sigma(0) = \sigma_{\rm I}, \;\;\;\;\;
 \beta(0) = \beta_{\rm I}, \nonumber\\
 \dot{a}(0) &= \sqrt{\frac{\hat{\rho}^{\rm tot}(0)}{3}}, 
 \;\;\;\;\;
 \dot{b}(0) = \dot{\sigma}(0) = 0, \label{ini_values}
\end{align}
where $b_{\rm I}$, $\sigma_{\rm I}$ and $\beta_{\rm I}$ are positive constants. 
The value of $\dot{a}(0)$ is determined by the constraint~(\ref{cstrt}).

\section{Transition from 5D to 3D spaces} 
\label{Approximations}

\subsection{Late-time behavior}
We focus on a situation that the standard 4D cosmology is realized at late times. 
Namely, the $S^2$ size modulus~$b$ and the dilaton~$\sigma$ are expected to be stabilized. 
In such a case, $\dot{B}=\dot{b}/b$ and $\dot{\sigma}$ become negligible, and (\ref{bg:EOM}) is reduced to 
\begin{align}
 \frac{3\dot{a}^2}{a^2}+\frac{1}{b^2}-\frac{e^\sigma}{8b^4}-V(\sigma)-\rho^{\rm rad} 
 &\simeq 0, \nonumber\\
 \frac{2\ddot{a}}{a}+\frac{\dot{a}^2}{a^2}+\frac{1}{b^2}-\frac{e^\sigma}{8b^4}-V(\sigma)+p^{\rm rad}_3 &\simeq 0, \nonumber\\
 \frac{3\ddot{a}}{a}+\frac{3\dot{a}^2}{a^2}+\frac{e^\sigma}{8b^4}-V(\sigma)+p^{\rm rad}_2 &\simeq 0, \nonumber\\
 \frac{e^\sigma}{8b^4}+\partial_\sigma V(\sigma) &\simeq 0. 
 \label{ap:bg:EOM}
\end{align}
Note that terms~$1/b^2$ and $-e^\sigma/(8b^4)$ come from the curvature of $S^2$ 
and the background flux of the U(1) gauge field, respectively. 
If they are cancelled with the potential term~$V(\sigma)$, we obtain the standard 4D Friedmann equations. 
This condition is written as
\begin{align}
 V_{\rm mdl}(b,\sigma) &\equiv -\frac{1}{b^2}+\frac{e^\sigma}{8b^4}+V(\sigma) \nonumber\\
 &= 2\brkt{e^{-\frac{\sigma}{2}}-\frac{e^{\sigma/2}}{4b^2}}^2+\frac{m^2}{2}\brkt{\sigma-\sigma_*}^2 = 0. 
 \label{V_mdl}
\end{align}
Namely, this is equivalent to the minimization condition of the moduli potential. 
From this condition, the stabilized moduli values are given by
\begin{align}
 \sigma &= \sigma_*, \;\;\;\;\;
 b = b_* = \frac{e^{\sigma_*/2}}{2}. 
 \label{stblz_value}
\end{align}
Then the Kaluza-Klein masses for a 6D massless field are given by  
\begin{align}
 m_{\rm KK}^{(l)} &= \frac{\sqrt{l(l+1)}}{b_*}, 
\end{align}
where $l=0,1,2,\cdots$. 
In particular, $m_{\rm KK}^{(1)}=\sqrt{2}/b_*$ is regarded as the cutoff energy scale of the 4D EFT. 

After the moduli are stabilized, the first two equations in (\ref{ap:bg:EOM}) are reduced to the 4D Friedmann equations.\footnote{
Precisely, we should note that the energy density and the pressure in the 4D spacetime are given by
$\rho_{\rm 4D}\equiv \rho^{\rm rad}{\cal V}_2$ and $P_{\rm 4D}\equiv p^{\rm rad}_3{\cal V}_2$, respectively, 
where ${\cal V}_2\equiv 4\pi b_*^2$ is the stabilized $S^2$ volume. 
Besides, the 4D Planck mass is $M_{\rm Pl}\equiv\sqrt{{\cal V}_2}$ since we have chosen the 6D Planck unit. 
Thus (\ref{4DFDeq}) should be expressed as 
\begin{align}
 \frac{\dot{a}^2}{a^2} &\simeq \frac{\rho_{\rm 4D}}{3M_{\rm Pl}^2}, \;\;\;\;\;
 \frac{\ddot{a}}{a} \simeq -\frac{1}{6M_{\rm Pl}^2}\brkt{\rho_{\rm 4D}+3P_{\rm 4D}}, 
\end{align}
as the 4D Friedmann equations. 
} 
\begin{align}
 \frac{\dot{a}^2}{a^2} \simeq \frac{\rho^{\rm rad}}{3}, \;\;\;\;\;
 \frac{\ddot{a}}{a} \simeq -\frac{1}{2}\brkt{\frac{\dot{a}^2}{a^2}+p_3^{\rm rad}}
 \simeq -\frac{1}{6}\brkt{\rho^{\rm rad}+3p^{\rm rad}_3}. 
 \label{4DFDeq}
\end{align}
The third equation becomes
\begin{align}
 \frac{3\ddot{a}}{a}+\frac{3\dot{a}^2}{a^2}+p_2^{\rm rad} &\simeq 0. 
 \label{ap:2Dcomp:Eq}
\end{align}
Using (\ref{4DFDeq}), the LHS is rewritten as
\begin{align}
  \frac{3\ddot{a}}{a}+\frac{3\dot{a}^2}{a^2}+p_2^{\rm rad} &\simeq -\frac{1}{2}\brkt{\rho^{\rm rad}+3p_3^{\rm rad}}+\rho^{\rm rad}+p_2^{\rm rad} 
  \nonumber\\ 
 &= \frac{1}{2}\brkt{\rho^{\rm rad}-3p_3^{\rm rad}}+p_2^{\rm rad} = 2p^{\rm rad}_2. 
\end{align}
At the last step, we have used the relation~(\ref{rel:rhop}). 
From (\ref{expr:rhop}) and Fig.~\ref{profile:Qs} in Appendix~\ref{TDquantities}, we find that $p_2^{\rm rad}\simeq 0$ at late times. 
Thus, (\ref{ap:2Dcomp:Eq}) holds at late times. 
In this case, the relation~(\ref{rel:rhop}) becomes $\rho^{\rm rad}\simeq 3p^{\rm rad}_3$, 
which indicates that the 4D universe is radiation-dominated. 
The last equation in (\ref{ap:bg:EOM}) holds trivially with the values in (\ref{stblz_value}). 

Since the functions~$Q_i(x)$ ($i=1,2,3$) are damped to zero for large $x$ (see Fig.~\ref{profile:Qs}), 
the evolution equation for the inverse temperature~(\ref{evolv_eq2}) becomes
\begin{align}
 \frac{\dot{\beta}}{\beta} &\simeq \dot{A} = \frac{\dot{a}}{a}, 
\end{align}
at late times. 
This indicates that $\beta\propto a$, which agrees with the relation in the 4D radiation-dominated era. 
 
In the following, we focus on the spacetime evolution during the moduli stabilization process before it settles into the above 4D evolution.

\subsection{Relation between {\boldmath $a$} and {\boldmath $\beta$}} \label{rel:a-bt}
As we have seen, our model realizes the 4D radiation-dominated universe at late times 
if the moduli are stabilized to the values in (\ref{stblz_value}). 
However, as shown in our previous work~\cite{Otsuka:2022rpx}, 
the behaviors of the 3D scale factor~$a$ and the (inverse) temperature~$\beta$ at early times can be different from 
those of the 4D universe. 

\begin{figure}[t]
  \begin{center}
    \includegraphics[scale=0.65]{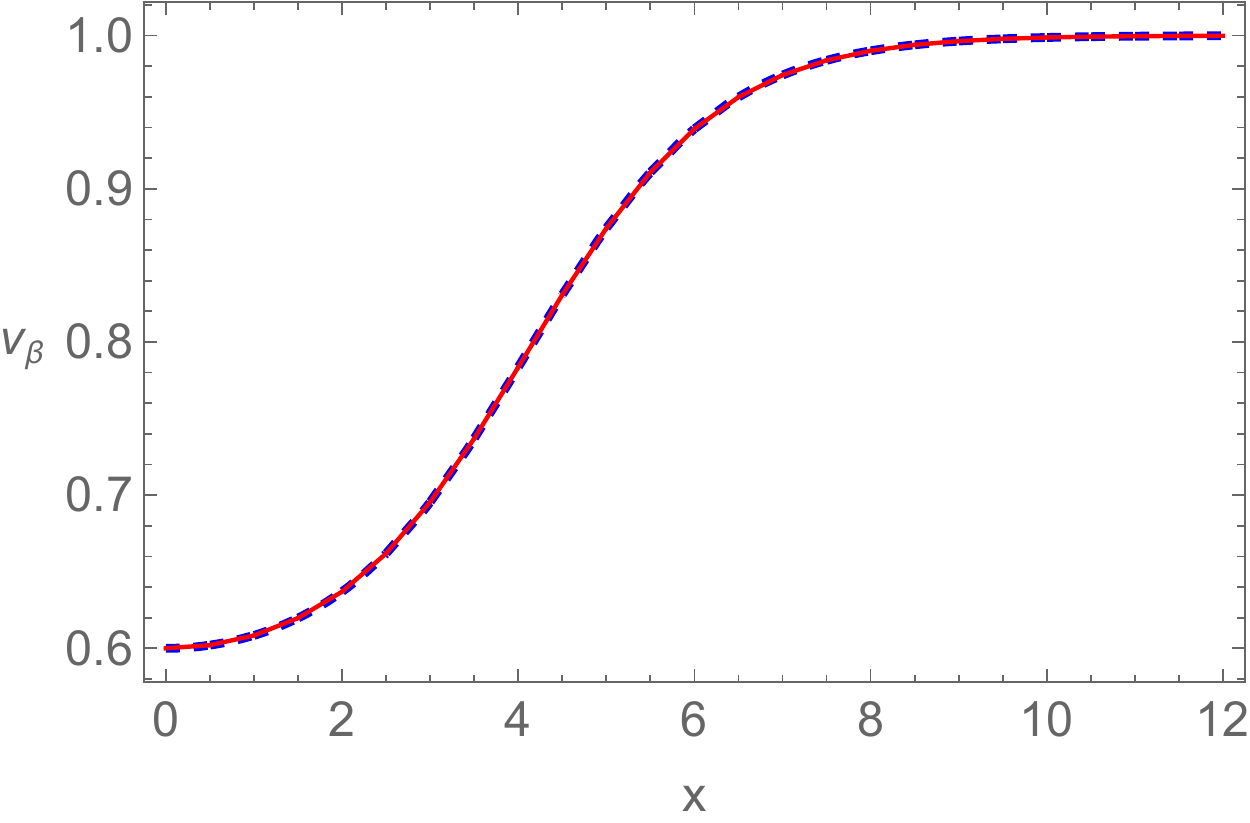} 
  \end{center}
\caption{The profiles of $v_\beta(x)$ (blue dashed) and $v_\beta^{\rm ap}(x)$ (red solid) in the case of $\Delta=0.5$. }
    \label{profile:vbt}
\end{figure}

We can see that $\dot{A}\gg |\dot{B}|$ during the moduli stabilization process. 
Thus (\ref{evolv_eq2}) is approximated as 
\begin{align}
 \frac{\dot{\beta}}{\beta} &= v_\beta\brkt{\frac{\beta}{b_*}}\frac{\dot{a}}{a}, 
 \label{rel:dotbt-dota}
\end{align} 
where
\begin{align}
 v_\beta(x) &= \frac{24\zeta(4)+12Q_1(x)+3Q_2(x)}{24\zeta(4)+12Q_1(x)+5Q_2(x)+Q_3(x)}. 
\end{align}
The profile of $v_\beta(x)$ is shown in Fig.~\ref{profile:vbt} (blue dashed line). 
As the plot shows, $v_\beta(x)\simeq 1$ for $x\geq 10$.  
For smaller values of $x$, $v_\beta(x)$ has a nontrivial profile. 
We approximate it by a piecewise-linear function for $0<x\leq 10$. 
Namely, we divide this interval into $(J+1)$ small intervals~$x_j<x\leq x_{j+1}$ ($j=0,1,2,\cdots,J$), 
where $x_j\equiv j\Delta$ ($\Delta\equiv 10/J$), 
and define the approximated function: 
\begin{align}
 v_\beta^{\rm ap}(x) &\equiv \begin{cases} c_1^{(j)}x+c_2^{(j)} & (x_j < x \leq x_{j+1} \leq x_J=10) \\
 1 & (x>10) \end{cases},  
\end{align}
where the constants~$c_1^{(j)}$ and $c_2^{(j)}$ are determined 
so that $v_\beta^{\rm ap}(x_j)=v_\beta(x_j)$ and $v_\beta^{\rm ap}(x_{j+1})=v_\beta(x_{j+1})$, i.e., 
\begin{align}
 c_1^{(j)} &\equiv \frac{v_\beta(x_{j+1})-v_\beta(x_j)}{\Delta}, \;\;\;\;\;
 c_2^{(j)} \equiv \frac{x_{j+1}v_\beta(x_j)-x_jv_\beta(x_{j+1})}{\Delta}. 
\end{align}
In the following, we choose $\Delta=0.5$ (or $J=20$). 
As shown in Fig.~\ref{profile:vbt}, $v_\beta^{\rm ap}(x)$ well approximates $v_\beta(x)$ in this case. 
In Sec.~\ref{Mdl_osc}, we will set a reference time~$t_{\rm ref}$ 
so that the evolution equations for $\tilde{B}\equiv B-B_*$ and $\tilde{\sigma}\equiv \sigma-\sigma_*$ can be approximated as 
homogeneous second-order linear differential equations for $t\geq t_{\rm ref}$. 
For this reference time, we define the integer~$k$ so that $x_k <\beta(t_{\rm ref})/b_* \leq x_{k+1}$. 
Then, by solving (\ref{rel:dotbt-dota}), the 3D scale factor~$a$ is expressed as 
\begin{align}
 a(x) &= \begin{cases} \displaystyle \frac{a_{\rm ref}K_k(x_{\rm ref})}{K_k(x)} & 
 \brkt{x_{\rm ref}\leq x\leq x_{k+1}} \\
 \displaystyle a_{\rm ref}K_k(x_{\rm ref})K_{k+1}(x_{k+1})\cdots K_j(x_j)\frac{1}{K_j(x)} & \brkt{x_j<x\leq x_{j+1}\leq x_J=10} \\
 \displaystyle \frac{a_{\rm ref}x}{10}K_k(x_{\rm ref})K_{k+1}(x_{k+1})\cdots K_{J-1}(x_{J-1}) & \brkt{x>10} \end{cases}, 
 \label{Expr:a-x}
\end{align}
where $x\equiv \beta/b_*$, $a_{\rm ref}\equiv a(t_{\rm ref})$, $x_{\rm ref}\equiv\beta(t_{\rm ref})/b_*$, 
and the function~$K_j(x)$ is defined by (\ref{def:K_j}). 
The detailed derivation is listed in Appendix~\ref{derivation_a}. 

In the case that $x_{\rm ref}\leq x_1=\Delta$, i.e., $k=0$, the expression of $a(\beta)$ becomes simple.  
Since 
\begin{align}
 0< c_1^{(0)} &\ll 1, \;\;\;\;\;
 c_2^{(0)} \simeq \frac{3}{5}, 
\end{align}
$K_0(x)$ is approximated as
\begin{align}
 K_0(x) &\simeq \brkt{\frac{x_1}{x}}^{5/3}. 
\end{align}
From (\ref{expr:a:k}), we have
\begin{align}
 a(x) &= \frac{a_{\rm ref}K_0(x_{\rm ref})}{K_0(x)} = a_{\rm ref}\brkt{\frac{x}{x_{\rm ref}}}^{5/3}. 
\end{align}
Hence, the relation between $a$ and $\beta$ in this case is obtained as  
\begin{align}
 \beta(a) &\simeq \beta_{\rm ref}\brkt{\frac{a}{a_{\rm ref}}}^{3/5}. 
 \label{beta-a:et}
\end{align}
for $a_{\rm ref}\leq a\leq a_1$.

\subsection{Radiation energy density}
\begin{figure}[t]
  \begin{center}
    \includegraphics[scale=0.65]{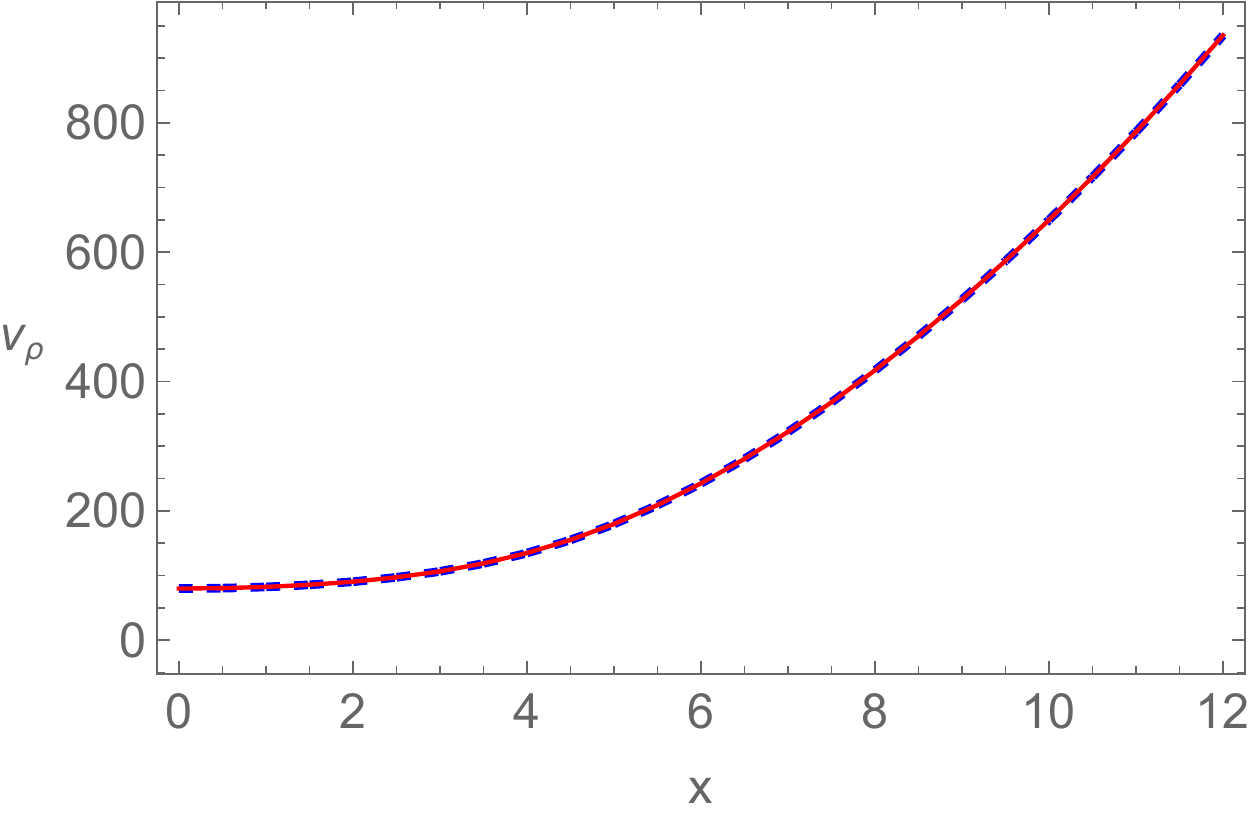} 
  \end{center}
\caption{The profiles of $v_\rho(x)$ (blue dashed) and $v_\rho^{\rm ap}(x)$ (red solid) in the case of $\Delta=0.5$. }
    \label{profile:vrho}
\end{figure}

During the moduli-stabilization process, the modulus~$b$ takes values around $b_*$. 
Thus, the radiation energy density~(\ref{expr:rhop}) is expressed as
\begin{align}
  \rho^{\rm rad} &\simeq \frac{g_{\rm dof}}{8\pi^3\beta^6}v_\rho\brkt{\frac{\beta}{b_*}}, 
  \label{ap:rho^rad}
\end{align}
where
\begin{align}
 v_\rho(x) &\equiv x^2\brc{6\zeta(4)+3Q_1(x)+Q_2(x)}. 
 \label{def:v_rho}
\end{align}
Note that the contributions of $Q_1(x)$ and $Q_2(x)$ are exponentially suppressed and negligible for $x>10$. 
Thus, just like the treatment for $v_\beta(x)$ in (\ref{rel:dotbt-dota}), we approximate $v_\rho(x)$ by 
\begin{align}
 v_\rho^{\rm ap}(x) &\equiv \begin{cases} d_1^{(j)}x+d_2^{(j)} & (x_j<x\leq x_{j+1}\leq 10) \\
 6\zeta(4)x^2 & (x>10) \end{cases}, 
 \label{ap:v_rho}
\end{align}
where
\begin{align}
 d_1^{(j)} &\equiv \frac{v_\rho(x_{j+1})-v_\rho(x_j)}{\Delta}, \;\;\;\;\;
 d_2^{(j)} \equiv \frac{x_{j+1}v_\rho(x_j)-x_j v_\rho(x_{j+1})}{\Delta}. 
\end{align}
As shown in Fig.~\ref{profile:vrho}, $v_\rho^{\rm ap}(x)$ well approximates $v_\rho(x)$. 

In particular, when $\beta<b_*\Delta$, (\ref{beta-a:et}) can be used and 
\begin{align}
 v_\rho\brkt{\frac{\beta}{b_*}} &\simeq 80, 
\end{align}
since 
\begin{align}
 d_1^{(0)} &= 1.27 \ll  d_2^{(0)} \simeq 80. 
\end{align}
Hence, in this case, $\rho^{\rm rad}$ can be expressed as the following simple function of $a$: 
\begin{align}
 \rho^{\rm rad} &\simeq \frac{10g_{\rm dof}}{\pi^3\beta^6} \simeq \frac{10g_{\rm dof}}{\pi^3\beta_{\rm ref}^6}
 \brkt{\frac{a}{a_{\rm ref}}}^{-\frac{18}{5}}. 
 \label{ap:rho^rad:et}
\end{align}
At the second step, we have used the relation~(\ref{beta-a:et}). 
This can also be derived directly from (\ref{expr:rhop}). 
In fact, when $\beta<b_*\Delta$, the functions~$Q_i$ ($i=1,2,3$) are approximated as
\begin{align}
 Q_1\brkt{\frac{\beta}{b_*}} &\simeq \frac{16b_*^2}{\beta^2}, \;\;\;\;\;
 Q_2\brkt{\frac{\beta}{b_*}} \simeq \frac{32b_*^2}{\beta^2}, \;\;\;\;\;
 Q_3\brkt{\frac{\beta}{b_*}} \simeq \frac{128b_*^2}{\beta^2}, 
\end{align}
which lead to
\begin{align}
 \rho^{\rm rad} &\simeq \frac{g_{\rm dof}}{8\pi^3b_*^2\beta^4}\brc{6\zeta(4)+\frac{48b_*^2}{\beta^2}+\frac{32b_*^2}{\beta^2}} 
 \simeq \frac{10g_{\rm dof}}{\pi^3\beta^6} \simeq \frac{10g_{\rm dof}}{\pi^3\beta_{\rm ref}^6}\brkt{\frac{a}{a_{\rm ref}}}^{-\frac{18}{5}}, \nonumber\\
 p^{\rm rad}_3 &\simeq \frac{g_{\rm dof}}{8\pi^3b_*^2\beta^4}\brc{2\zeta(4)+\frac{16b_*^2}{\beta^2}}
 \simeq \frac{2g_{\rm dof}}{\pi^3\beta^6} \simeq \frac{2g_{\rm dof}}{\pi^3\beta_{\rm ref}^6}\brkt{\frac{a}{a_{\rm ref}}}^{-\frac{18}{5}}, \nonumber\\
 p^{\rm rad}_2 &\simeq \frac{g_{\rm dof}}{16\pi^3b_*^2\beta^4}\cdot\frac{32b_*^2}{\beta^2} = \frac{2g_{\rm dof}}{\pi^3\beta^6}
 \simeq \frac{2g_{\rm dof}}{\pi^3\beta_{\rm ref}^6}\brkt{\frac{a}{a_{\rm ref}}}^{-\frac{18}{5}}. 
 \label{ap:expr:rhop}
\end{align}
Note that the pressure becomes isotropic~$p_3^{\rm rad}\simeq p_2^{\rm rad}$ in this case 
because the radiation does not feel the size of the compact space~$S^2$. 
In fact, the above pressures are (almost) independent of $b_*$.

\subsection{Moduli oscillation} \label{Mdl_osc}
Here we discuss the moduli oscillation around their stabilized values in (\ref{stblz_value}). 
We divide $B$ and $\sigma$ as
\begin{align}
 B &= B_*+\tl{B}, \;\;\;\;\;
 \sigma = \sigma_*+\tl{\sigma}, 
\end{align}
where
\begin{align}
 B_* &\equiv \ln b_* = \frac{\sigma_*}{2}-\ln 2. 
\end{align}
Then the evolution equations in (\ref{evolv_eq1}) and (\ref{evolv_eq:B}) are written as
\begin{align}
 \ddot{A} &= -\frac{9}{4}\dot{A}^2-\frac{1}{2}\dot{A}\dot{\tl{B}}+2e^{-\sigma_*}\brkt{-\tl{\sigma}+2\tl{B}}
 +\frac{p_3^{\rm rad}-2p_2^{\rm rad}}{4}+\cdots, \nonumber\\
 \ddot{\tl{B}} &= -3\dot{A}\dot{\tilde{B}}-2e^{-\sigma_*}\brkt{-\tilde{\sigma}+2\tilde{B}}+p_2^{\rm rad}+\cdots, \nonumber\\
 \ddot{\tl{\sigma}} &= -3\dot{A}\dot{\tl{\sigma}}+4e^{-\sigma_*}\brkt{-\tl{\sigma}+2\tl{B}}-m^2\tl{\sigma}+\cdots, 
 \label{ap:EOM}
\end{align}
and the constraint~(\ref{cstrt}) becomes 
\begin{align}
 3\dot{A}^2+\dot{\tl{B}}^2+6\dot{A}\dot{\tl{B}}-\frac{1}{2}\dot{\tl{\sigma}}^2 
 &= 2e^{-\sigma_*}\brkt{\tl{\sigma}-2\tl{B}}^2+\frac{m^2}{2}\tl{\sigma}^2+\rho^{\rm rad}+\cdots, 
 \label{ap:cstrt}
\end{align}
where the ellipses denote higher-order terms in $\tl{B}$ or $\tl{\sigma}$. 

\begin{figure}[t]
  \begin{center}
    \includegraphics[scale=0.6]{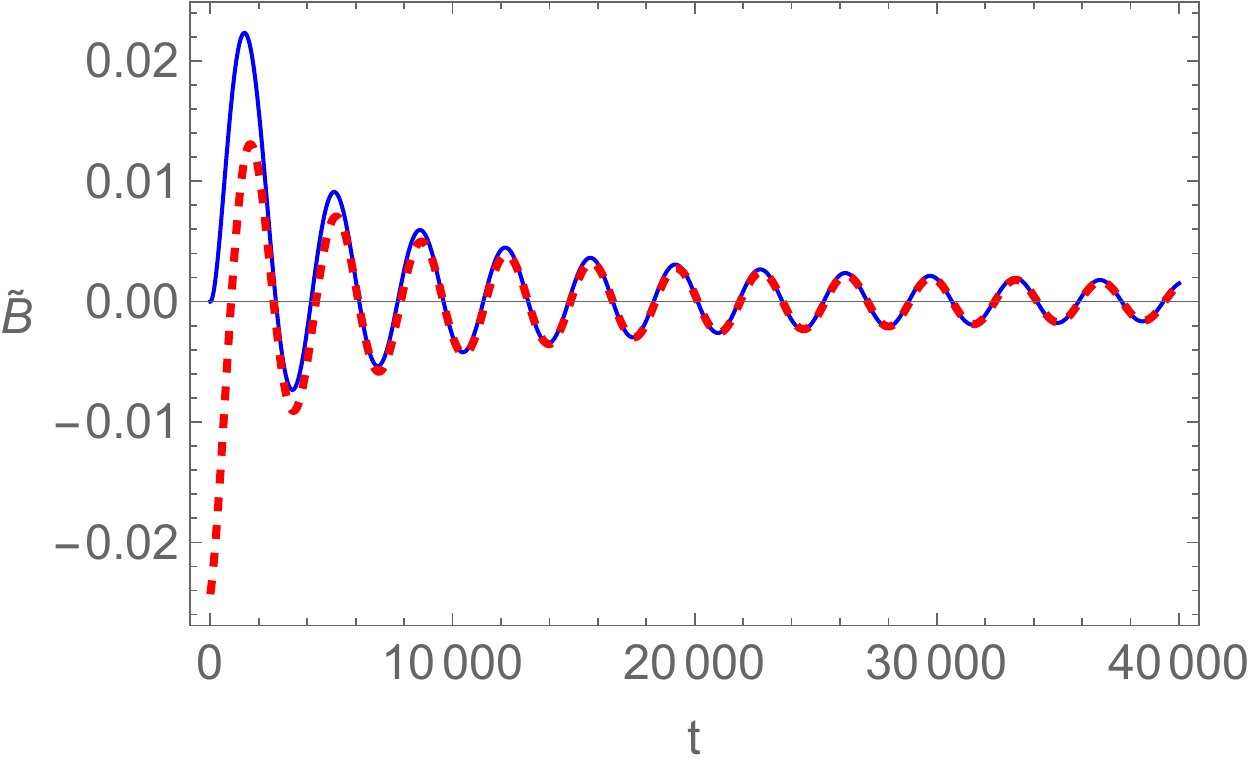} \;\;
    \includegraphics[scale=0.6]{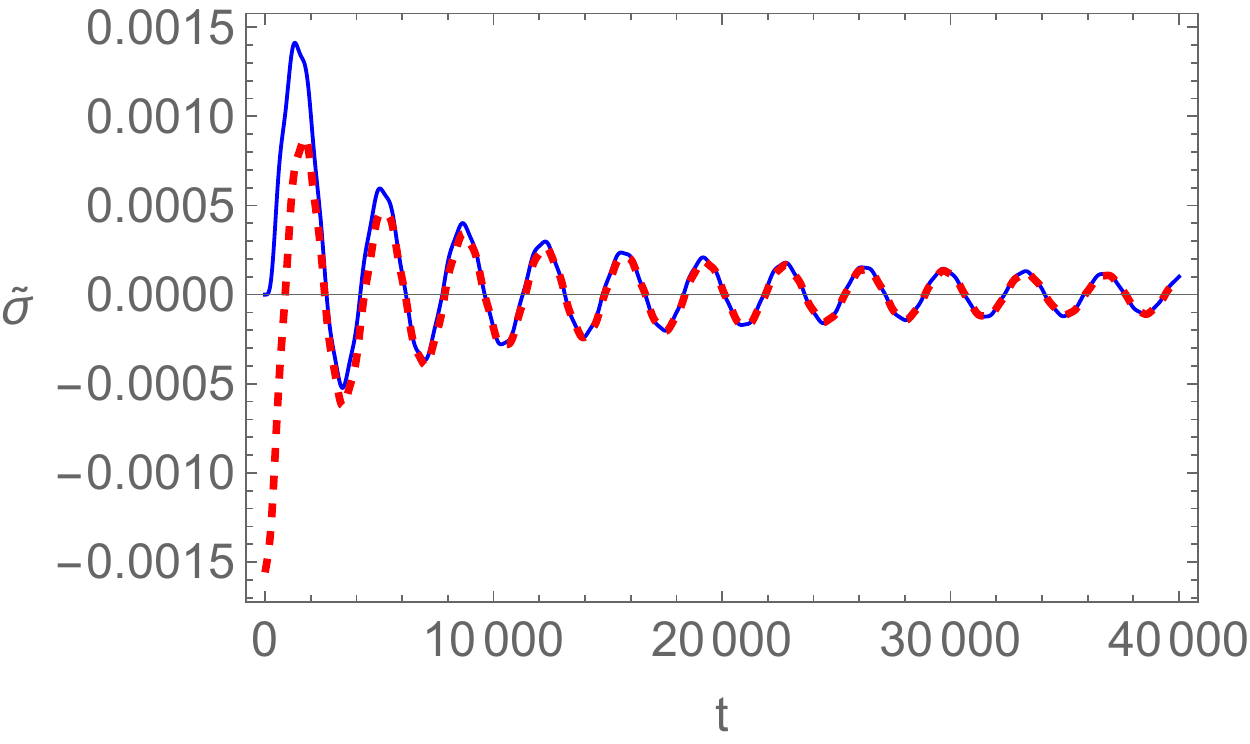} 
  \end{center}
\caption{The profiles of $\tilde{B}(t)$ (left) and $\tilde{\sigma}(t)$ (right) in the case of $t_{\rm ref}=10000$. 
The blue solid lines are the full numerical solutions, and the red dashed lines are the approximate solutions in (\ref{ap:B-sgm}), 
where $\bar{A}(t)$ is computed by the method in Sec.~\ref{transit:53}. 
The parameters are chosen as (\ref{parameter_choice:1}).  }
    \label{profiles:tlBsgm}
\end{figure}

We introduce the reference time~$t_{\rm ref}$ so that 
the last term in the second equation of (\ref{ap:EOM}) becomes negligible for $t\geq t_{\rm ref}$. 
In the following, we choose it as $t_{\rm ref}=10000$ (in the 6D Planck unit). 
Then, from the first equation in (\ref{ap:EOM}) and (\ref{ap:cstrt}), we find that 
\begin{align}
 \ddot{A} &= {\cal O}(\tl{B},\tl{\sigma})
\end{align}
Thus, the second and the third equations in (\ref{ap:EOM}) can be rewritten as
\begin{align}
 \begin{pmatrix} 2\ddot{\hat{B}} \\ \ddot{\hat{\sigma}} \end{pmatrix} 
 &= -\begin{pmatrix} 4e^{-\sigma_*} & -4e^{-\sigma_*} \\-4e^{-\sigma_*} & 4e^{-\sigma_*}+m^2 \end{pmatrix}
 \begin{pmatrix} 2\hat{B} \\ \hat{\sigma} \end{pmatrix}+\cdots, 
\end{align}
where
\begin{align}
 \hat{B} &\equiv e^{\frac{3}{2}A}\tl{B}, \;\;\;\;\;
 \hat{\sigma} \equiv e^{\frac{3}{2}A}\tl{\sigma}. 
\end{align}
Diagonalizing this, we have
\begin{align}
 \begin{pmatrix} \ddot{\varphi}_1 \\ \ddot{\varphi}_2 \end{pmatrix} 
 &= -\begin{pmatrix} \lambda_1 & \\ & \lambda_2 \end{pmatrix}
 \begin{pmatrix} \varphi_1 \\ \varphi_2 \end{pmatrix}+\cdots, 
 \label{eq:vph12}
\end{align}
where 
\begin{align}
 \lambda_1 &\equiv \frac{1}{2}\brkt{8e^{-\sigma_*}+m^2-\sqrt{64e^{-2\sigma_*}+m^4}} 
 = \frac{1}{2}\brkt{m_{\rm KK}^{(1)2}+m^2-\sqrt{m_{\rm KK}^{(1)4}+m^4}}, \nonumber\\
 \lambda_2 &\equiv  \frac{1}{2}\brkt{8e^{-\sigma_*}+m^2+\sqrt{64e^{-2\sigma_*}+m^4}} 
 = \frac{1}{2}\brkt{m_{\rm KK}^{(1)2}+m^2+\sqrt{m_{\rm KK}^{(1)4}+m^4}}, 
 \label{eigenvalues}
\end{align}
and
\begin{align}
 \varphi_1 &\equiv \cos\theta\cdot 2\hat{B}+\sin\theta\cdot \hat{\sigma}, \nonumber\\
 \varphi_2 &\equiv -\sin\theta\cdot 2\hat{B}+\cos\theta\cdot\hat{\sigma}, 
 \label{def:varphi}
\end{align}
with
\begin{align}
 \theta &\equiv \tan^{-1}\frac{8e^{-\sigma_*}}{m^2+\sqrt{64e^{-2\sigma_*}+m^4}}
 = \tan^{-1}\frac{m_{\rm KK}^{(1)2}}{m^2+\sqrt{m_{\rm KK}^{(1)4}+m^4}}.  \label{def:theta}
\end{align}
Solving (\ref{eq:vph12}), we have
\begin{align}
 \varphi_1(t) &= \frac{\chi_{10}}{\sqrt{\lambda_1}}\sin\brkt{\sqrt{\lambda_1}(t-t_{\rm ref})}
 +\varphi_{10}\cos\brkt{\sqrt{\lambda_1}(t-t_{\rm ref})}, \nonumber\\
 \varphi_2(t) &= \frac{\chi_{20}}{\sqrt{\lambda_2}}\sin\brkt{\sqrt{\lambda_2}(t-t_{\rm ref})}
 +\varphi_{20}\cos\brkt{\sqrt{\lambda_2}(t-t_{\rm ref})}, 
 \label{t-dep:vph}
\end{align}
where 
\begin{align}
 \varphi_{10} &\equiv e^{\frac{3}{2}A(t_{\rm ref})}\brc{2\cos\theta\tilde{B}(t_{\rm ref})+\sin\theta\tilde{\sigma}(t_{\rm ref})}, 
 \nonumber\\
 \varphi_{20} &\equiv e^{\frac{3}{2}A(t_{\rm ref})}\brc{-2\sin\theta\tilde{B}(t_{\rm ref})+\cos\theta\tilde{\sigma}(t_{\rm ref})}, 
 \nonumber\\
 \chi_{10} &\equiv e^{\frac{3}{2}A(t_{\rm ref})}\brc{2\cos\theta\dot{\tilde{B}}(t_{\rm ref})+\sin\theta\dot{\tilde{\sigma}}(t_{\rm ref})}
 +\frac{3}{2}\dot{A}(t_{\rm ref})\varphi_{10}, \nonumber\\
 \chi_{20} &\equiv e^{\frac{3}{2}A(t_{\rm ref})}\brc{-2\sin\theta\dot{\tilde{B}}(t_{\rm ref})+\cos\theta\dot{\tilde{\sigma}}(t_{\rm ref})}
 +\frac{3}{2}\dot{A}(t_{\rm ref})\varphi_{20}. 
\end{align}
In terms of $\varphi_1$ and $\varphi_2$, (the displacement of) the moduli~$\tilde{B}$ and $\tilde{\sigma}$ are expressed as 
\begin{align}
 \tilde{B}(t) &= \frac{e^{-\frac{3}{2}A(t)}}{2}\brc{\cos\theta\varphi_1(t)-\sin\theta\varphi_2(t)}, \nonumber\\
 \tilde{\sigma}(t) &= e^{-\frac{3}{2}A(t)}\brc{\sin\theta\varphi_1(t)+\cos\theta\varphi_2(t)}. 
 \label{Bsgm-vphs}
\end{align}
Since $A(t)\gg |\tilde{B}(t)|$, we can replace $A(t)$ in the exponents with $\bar{A}(t)\equiv A(t)+\tilde{B}(t)$. 
In the next subsection, we will show that $\bar{A}(t)$ has a nice property to calculate its approximate expression. 
Therefore, we express $\tilde{B}$ and $\tilde{\sigma}$ as 
\begin{align}
 \tilde{B}(t) &= \frac{e^{-\frac{3}{2}\bar{A}(t)}}{2}\brc{\cos\theta\varphi_1(t)-\sin\theta\varphi_2(t)}, \nonumber\\
 \tilde{\sigma}(t) &= e^{-\frac{3}{2}\bar{A}(t)}\brc{\sin\theta\varphi_1(t)+\cos\theta\varphi_2(t)}. 
 \label{ap:B-sgm}
\end{align}
Fig.~\ref{profiles:tlBsgm} shows the full numerical solutions and the approximate solutions in (\ref{ap:B-sgm}) for $\tilde{B}(t)$ and $\tilde{\sigma}(t)$. 
Here $\bar{A}(t)$ is computed by the method explained in Sec.~\ref{transit:53}. 
The parameters are chosen as
\begin{align}
 m &= 0.01, \;\;\;\;\;
 \sigma_* = 14 \;\; \brkt{ \Leftrightarrow m_{\rm KK}^{(1)} =0.00258}, \nonumber\\
 b_{\rm I} &= b_*, \;\;\;\;\; \sigma_{\rm I} = \sigma_*, \;\;\;\;\;
 \beta_{\rm I} = 20. 
 \label{parameter_choice:1}
\end{align}
We can see that the latters well agree with the formers for $t\geq t_{\rm ref}$. 

\subsection{Smoothing the oscillation of the 3D scale factor}
\begin{figure}[t]
  \begin{center}
    \includegraphics[scale=0.65]{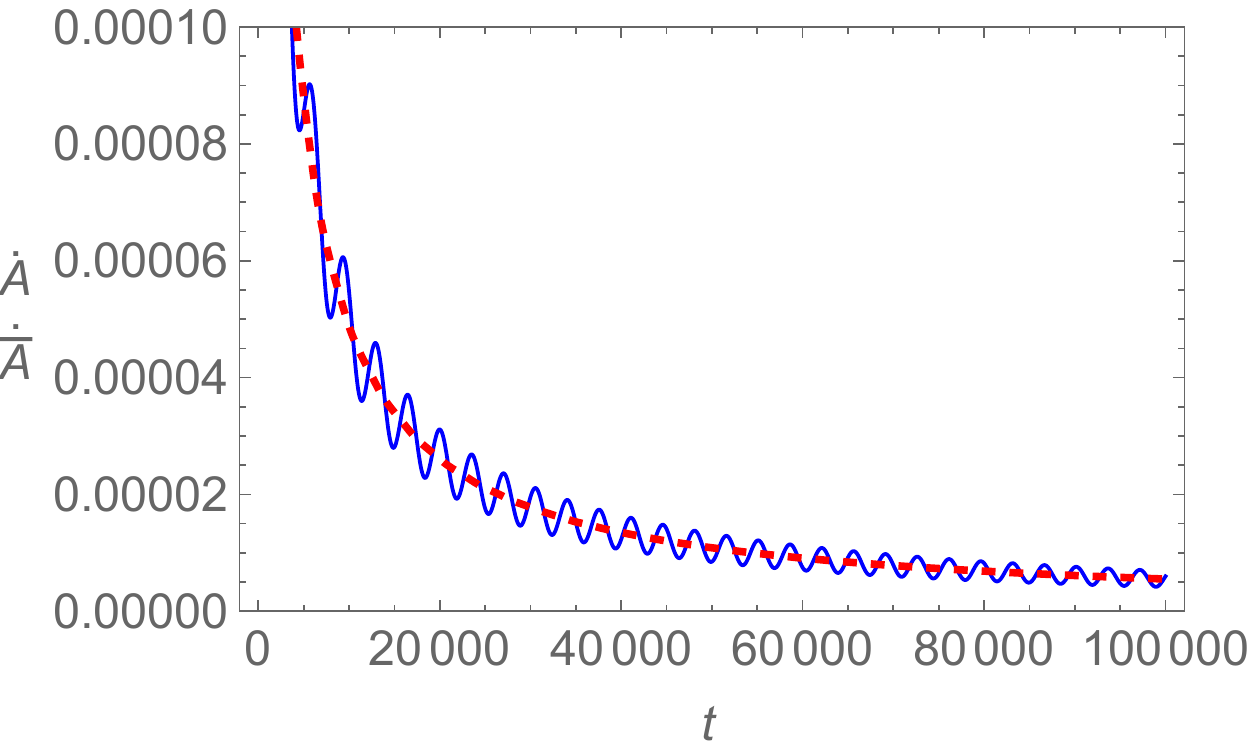} \;\;
    \includegraphics[scale=0.6]{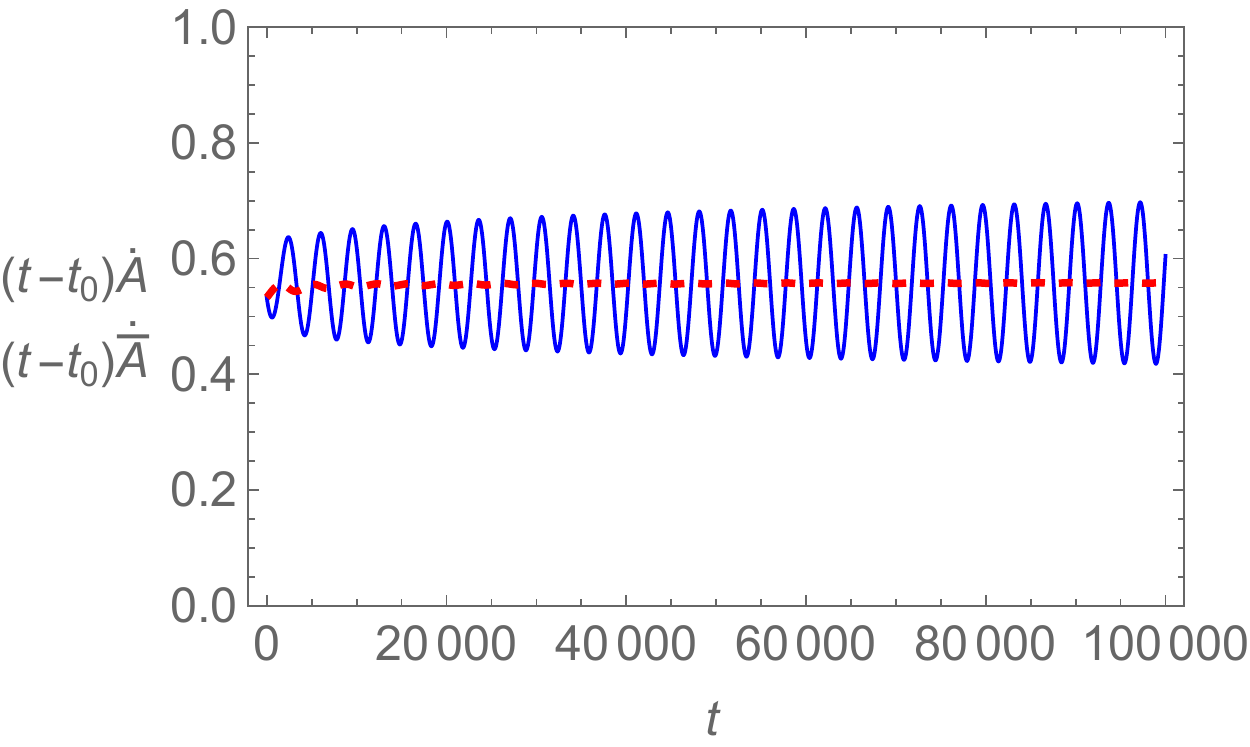}
  \end{center}
\caption{The profile of $\dot{A}(t)$ (blue solid) and $\dot{\bar{A}}(t)=\dot{A}(t)+\dot{B}(t)$ (red dashed) in the left plot. 
Those of $(t-t_0)\dot{A}(t)$ (blue solid) and $(t-t_0)\dot{\bar{A}}(t)$ (red dashed) in the right plot. 
The constant~$t_0$ and the reference time are chosen as $t_0=-1300$ and $t_{\rm ref}=10000$, respectively. 
The parameters are chosen as (\ref{parameter_choice:1}).  }
    \label{profile:dota}
\end{figure}

Now we derive an approximate expression for the 3D scale factor~$a(t)$ (or $A(t)$). 
For this purpose, it is convenient to define 
\begin{align}
 \bar{A}(t) &\equiv A(t)+\tilde{B}(t). 
\end{align}
Then, its evolution is determined by 
\begin{align}
 3\dot{\bar{A}}^2 &= 2\dot{B}^2+\frac{1}{2}\dot{\sigma}^2+2\brkt{e^{-\frac{\sigma}{2}}-\frac{e^{\frac{\sigma}{2}-2B}}{4}}^2
 +\frac{m^2}{2}\brkt{\sigma-\sigma_*}^2+\rho^{\rm rad} \nonumber\\
 &= 2\dot{B}^2+\frac{1}{2}\dot{\sigma}^2+\hat{\rho}^{\rm tot} \equiv \tilde{\rho}^{\rm tot}, 
 \label{def:tlrhotot}
\end{align}
which is obtained from the constraint~(\ref{ap:cstrt}). 
During the moduli stabilization process, the RHS is approximated as
\begin{align}
 3\dot{\bar{A}}^2 &\simeq 2\dot{\tilde{B}}^2+\frac{1}{2}\dot{\tilde{\sigma}}^2
 +2e^{-\sigma_*}\brkt{\tilde{\sigma}-2\tilde{B}}^2+\frac{m^2}{2}\tilde{\sigma}^2+\rho^{\rm rad}+\cdots 
 \nonumber\\
 &\simeq \frac{e^{-3\bar{A}}}{2}\brc{\cos\theta\dot{\varphi}_1-\sin\theta\dot{\varphi}_2
 -\frac{3}{2}\dot{\bar{A}}\brkt{\cos\theta\varphi_1-\sin\theta\varphi_2}}^2 \nonumber\\
 &\quad +\frac{e^{-3\bar{A}}}{2}\brc{\sin\theta\dot{\varphi}_1+\cos\theta\dot{\varphi}_2
 -\frac{3}{2}\dot{\bar{A}}\brkt{\sin\theta\varphi_1+\cos\theta\varphi_2}}^2 \nonumber\\
 &\quad +2e^{-3\bar{A}-\sigma_*}\brc{\brkt{\sin\theta-\cos\theta}\varphi_1+\brkt{\cos\theta+\sin\theta}\varphi_2}^2 \nonumber\\
 &\quad +\frac{m^2}{2}e^{-3\bar{A}}\brkt{\sin\theta\varphi_1+\cos\theta\varphi_2}^2+\rho^{\rm rad}
 +\cdots, \label{expr:cstrt:1}
\end{align}
where the ellipses denote higher order terms in $\tilde{B}$ or $\tilde{\sigma}$ (or $\varphi_{1,2}$). 
At the second step, we have used (\ref{ap:B-sgm}). 
The solutions in (\ref{t-dep:vph}) are rewritten as 
\begin{align}
 \varphi_1 &= \sqrt{\frac{\chi_{10}^2}{\lambda_1}+\varphi_{10}^2}\sin\brkt{\sqrt{\lambda_1}(t-t_{\rm ref})+\delta_1}, \nonumber\\
 \varphi_2 &= \sqrt{\frac{\chi_{20}^2}{\lambda_2}+\varphi_{20}^2}\sin\brkt{\sqrt{\lambda_2}(t-t_{\rm ref})+\delta_2}, 
 \label{ap:vphs}
\end{align}
where
\begin{align}
 \delta_1 &\equiv \tan^{-1}\brkt{\frac{\varphi_{10}}{\chi_{10}}\sqrt{\lambda_{10}}}, \;\;\;\;\;
 \delta_2 \equiv \tan^{-1}\brkt{\frac{\varphi_{20}}{\chi_{20}}\sqrt{\lambda_{20}}}. 
\end{align}
Here note 
\begin{align}
 \dot{\bar{A}}^2 \ll \lambda_1, \; \lambda_2 \ll 1, 
 \label{smallness_dotA}
\end{align} 
for $t\geq t_{\rm ref}$. 
Then, (\ref{expr:cstrt:1}) is simplified as
\begin{align}
 3\dot{\bar{A}} &\simeq \frac{e^{-3\bar{A}}}{2}\bigg[\dot{\varphi}_1^2+\dot{\varphi}_2^2
 +\brc{4e^{-\sigma_*}\brkt{\sin\theta-\cos\theta}^2+m^2\sin^2\theta}\varphi_1^2
 \nonumber\\
 &\quad \hspace{12mm}
 +\brc{8e^{-\sigma_*}\brkt{\sin^2\theta-\cos^2\theta}+2\sin\theta\cos\theta}\varphi_1\varphi_2 \nonumber\\
 &\quad \hspace{12mm}
 +\brc{4e^{-\sigma_*}\brkt{\cos\theta+\sin\theta}^2+m^2\cos^2\theta}\varphi_2^2\bigg]+\rho^{\rm rad}+\cdots \nonumber\\
 &= \frac{e^{-3\bar{A}}}{2}\brkt{\dot{\varphi}_1^2+\dot{\varphi}_2^2+\lambda_1\varphi_1^2+\lambda_2\varphi_2^2}+\rho^{\rm rad}+\cdots \nonumber\\
 &= C_1e^{-3\bar{A}}+\rho^{\rm rad}+\cdots, 
 \label{3barA^2}
\end{align}
where we have used that
\begin{align}
 4e^{-\sigma_*}\brkt{\sin\theta-\cos\theta}^2+m^2\sin^2\theta
 &= \lambda_1, \nonumber\\
 8e^{-\sigma_*}\brkt{\sin^2\theta-\cos^2\theta}+2\sin\theta\cos\theta
 &= 0, \nonumber\\
 4e^{-\sigma_*}\brkt{\cos\theta+\sin\theta}^2+m^2\cos^2\theta &= \lambda_2, 
\end{align}
and the constant~$C_1$ is defined as 
\begin{align}
 C_1 &\equiv \frac{1}{2}\brkt{\chi_{10}^2+\lambda_1\varphi_{10}^2+\chi_{20}^2+\lambda_2\varphi_{20}^2}. 
 \label{def:C1}
\end{align}
At the last step, we used (\ref{ap:vphs}). 
Recalling that $\rho^{\rm rad}$ is a function of $A$, the RHS of (\ref{3barA^2}) is almost a function of only $\bar{A}$, 
and its explicit $t$-dependence is negligible. 
In fact,  as we can see from Fig.~\ref{profile:dota}, the oscillating behavior of $\dot{A}$ is almost cancelled by adding $\dot{\tilde{B}}$. 
Hence we have~\footnote{
We are interested in only the expanding-universe solution ($\dot{\bar{A}}>0$). 
}
\begin{align}
 \dot{\bar{A}} &\simeq \sqrt{\frac{1}{3}\brkt{C_1e^{-3\bar{A}}+\rho^{\rm rad}}}. 
 \label{dotbarA}
\end{align}
From this, we obtain 
\begin{align}
 t-t_{\rm ref} &= \int_{\bar{A}(t_{\rm ref})}^{\bar{A}}dA\;\sqrt{\frac{3}{C_1e^{-3A}+\rho^{\rm rad}(A)}}. 
 \label{expr:t-a}
\end{align}
By taking the inverse function of this, 
the 3D scale factor~$a\simeq e^{\bar{A}}$ is obtained as a function of $t$. 

In the standard 4D cosmology, the 3D scale factor~$a(t)$ behaves as
\begin{align}
 a(t) &\propto (t-t_0)^p, \;\;\;\;\;\brkt{\mbox{$t_0$ : constant}}
 \label{behavior:a:4D}
\end{align}
where $p=1/2$ in the radiation-dominated era and $p=2/3$ in the matter-dominated era. 
If $a(t)$ behaves as (\ref{behavior:a:4D}), the power~$p$ is calculated as
\begin{align}
 p &\equiv (t-t_0)\frac{\dot{a}}{a} = (t-t_0)\dot{A}, 
 \label{def:power}
\end{align}
where the constant~$t_0$ is chosen so that $p$ becomes independent of $t$. 
In our setup, this quantity oscillates in time. 
Thus, it is convenient to use $\bar{A}$ instead of $A$ in order to define ``effective power''~$p$ 
(see the right plot of Fig.~\ref{profile:dota}). 
In fact, we can easily calculate this power in the radiation-dominated universe, 
and it turns out to be that $p=1/2$ in the 3D space while $p=5/9$ in the 5D space. 
In our setup, the universe eventually approaches a 3D space in which 
the moduli oscillation dominates the energy density. 
In such a case, $p$ takes 2/3, which is the same value as the matter-dominated 3D universe.

\subsection{Evolution of 3D space} \label{transit:53}
The expression~(\ref{expr:t-a}) can be rewritten as
\begin{align}
 t-t_{\rm ref} &\simeq \int_{x_{\rm ref}}^x d\tilde{x}\;\frac{d\bar{A}}{dx}(\tilde{x})
 \sqrt{\frac{3}{C_1e^{-3\bar{A}(\tilde{x})}+\rho^{\rm rad}(\tilde{x})}}, 
 \label{expr:t}
\end{align}
where~\footnote{
Precisely, $\rho^{\rm rad}(x)$ should be written as $\rho^{\rm rad}(\bar{A}(x))$. 
}
\begin{align}
 \rho^{\rm rad}(x) &= \frac{g_{\rm dof}}{8\pi^3b_*^6x^6}v_\rho^{\rm ap}(x). 
 \label{rhorad:ap}
\end{align}
From (\ref{Ax}), $d\bar{A}/dx$ is given by 
\begin{align}
 \frac{d\bar{A}}{dx}(x) &= \begin{cases} \displaystyle \frac{1}{c_2^{(j)}}\brkt{\frac{1}{x}-\frac{c_1^{(j)}}{c_1^{(j)}x+c_2^{(j)}}} 
 & (x_j<x\leq x_{j+1}\leq 10) \\
 \displaystyle x^{-1} & (x>10) \end{cases}, 
\end{align}
and from (\ref{expr:a:k})-(\ref{expr:a:2}), $e^{-3\bar{A}(x)}$ is given by
\begin{align}
 e^{-3\bar{A}(x)} &= \begin{cases} \displaystyle \brkt{\frac{K_k(x)}{a_{\rm ref}K_k(x_{\rm ref})}}^3 & (x_{\rm ref}\leq x\leq x_{k+1}) \\
 \displaystyle \brkt{\frac{K_j(x)}{a_{\rm ref}K_k(x_{\rm ref})K_{k+1}(x_{k+1})\cdots K_j(x_j)}}^3 & (x_j<x\leq x_{j+1}) \\
 \displaystyle \brkt{\frac{a_{\rm ref}x}{10}K_k(x_{\rm ref})K_{k+1}(x_{k+1})\cdots K_{J-1}(x_{J-1})}^{-3} & (x>10) 
 \end{cases}. 
 \label{e3A:ap}
\end{align}
From these expressions, we can numerically compute the 3D scale factor~$a\simeq e^{\bar{A}}$ at an arbitrary time 
through the auxiliary variable~$x$, which is the (normalized) inverse temperature. 
This approximation makes it easier to compute $a$ at a much later time than $t=t_{\rm ref}$, 
which cannot practically be obtained by the full numerical computation.

\begin{figure}[t]
  \begin{center}
    \includegraphics[scale=0.65]{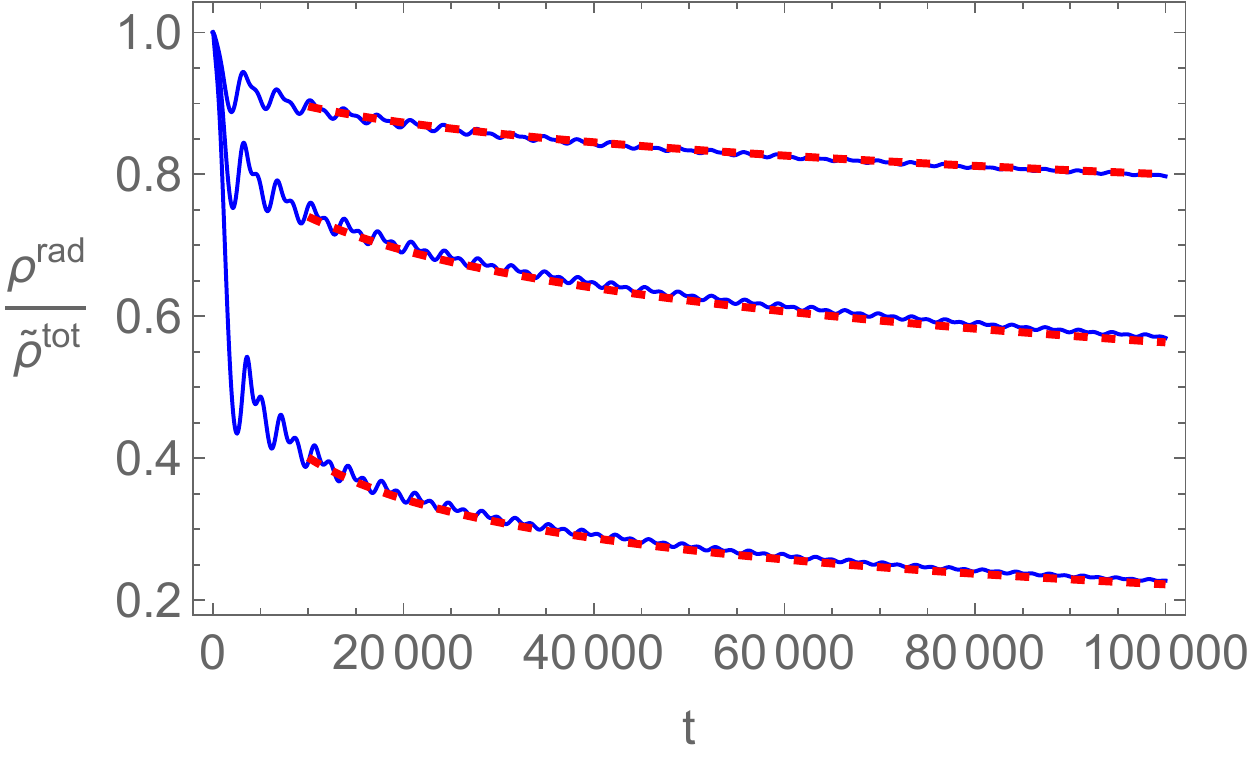} \;
    \includegraphics[scale=0.65]{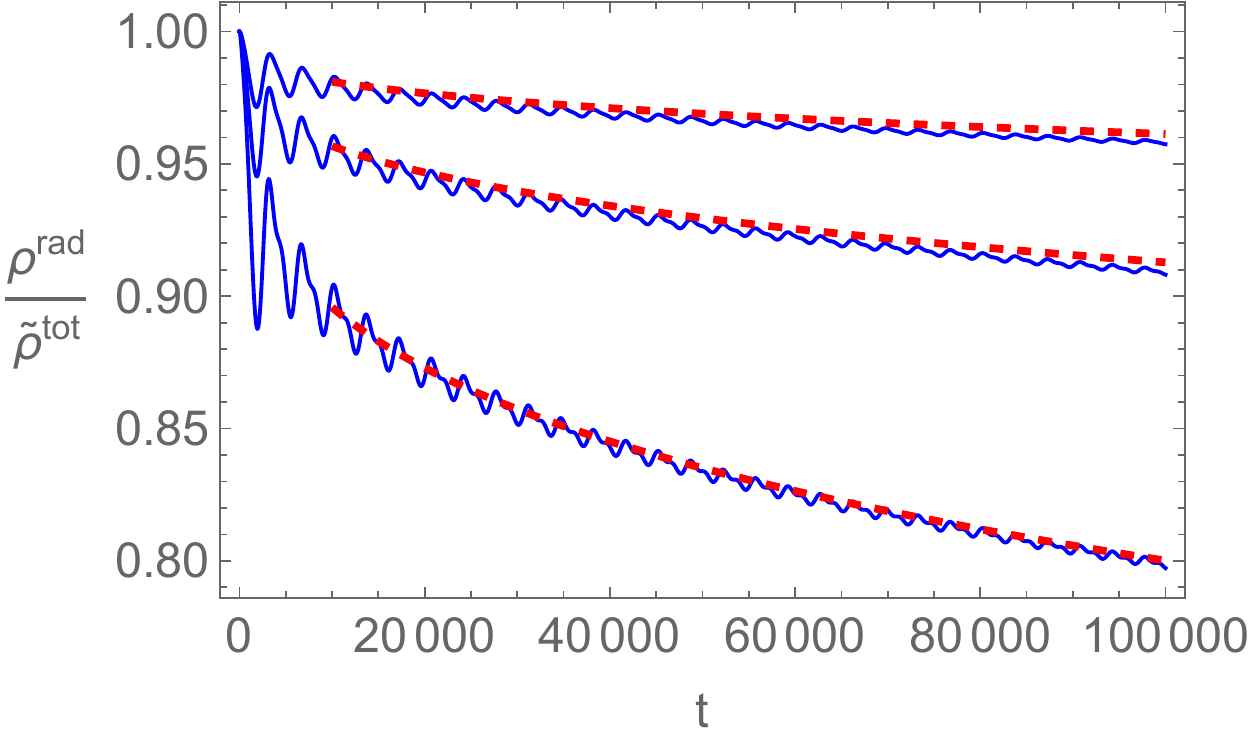}
  \end{center}
\caption{The time evolution of the ratio~$\rho^{\rm rad}/\tilde{\rho}^{\rm tot}$. 
The blue solid and the red dashed lines represent the full numerical calculations and the approximate ones, respectively. 
The lines from bottom to top correspond to $\beta_{\rm I}=10,12.5,15$ (left plot) and $\beta_{\rm I}=15,17.5,20$ (right plot).  
The other parameters are chosen as (\ref{parameter_choice:1}). 
The reference time is chosen as $t_{\rm ref}=10000$. }
    \label{rhorto1}
\end{figure}

Here we check the validity of these approximations by comparing with the results of the full numerical computation. 
For $t\geq t_{\rm ref}$, the total energy density~$\tilde{\rho}^{\rm tot}$ is expressed as 
\begin{align}
 \tilde{\rho}^{\rm tot} &\equiv C_1e^{-3\bar{A}}+\rho^{\rm rad}(\bar{A}). 
\end{align}
Fig.~\ref{rhorto1} shows the ratio of the radiation energy density to the total energy density~$\rho^{\rm rad}/\tilde{\rho}^{\rm tot}$ 
as a function of $t$. 
The blue solid lines represent the full numerical results and the red dashed lines represent the approximate ones 
obtained by using (\ref{rhorad:ap}) and (\ref{e3A:ap}). 
The constant~$C_1$ defined in (\ref{def:C1}) is determined by numerically calculating the time evolution of the moduli up to $t=t_{\rm ref}$. 
We can see that our approximation well reproduces the full numerical results. 

\begin{figure}[t]
  \begin{center}
    \includegraphics[scale=0.62]{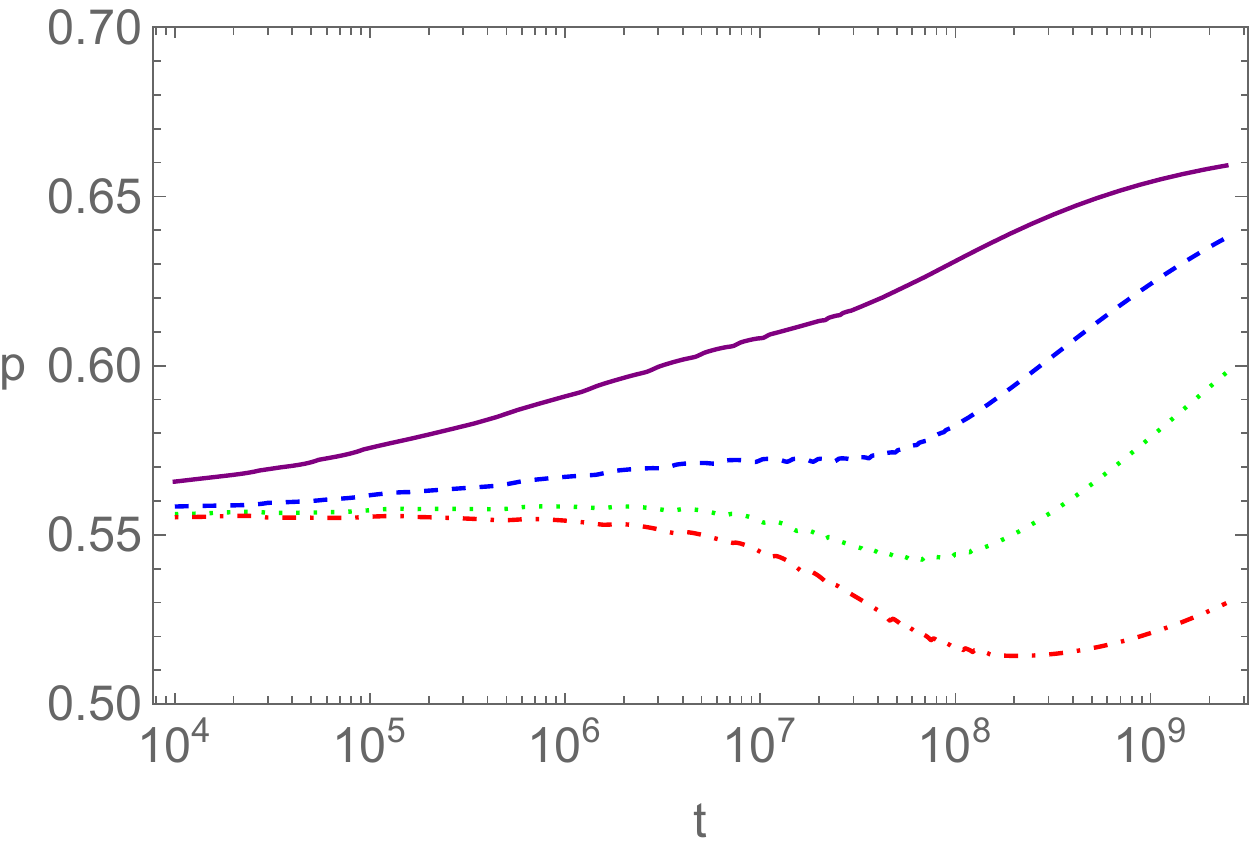} \;\;
    \includegraphics[scale=0.65]{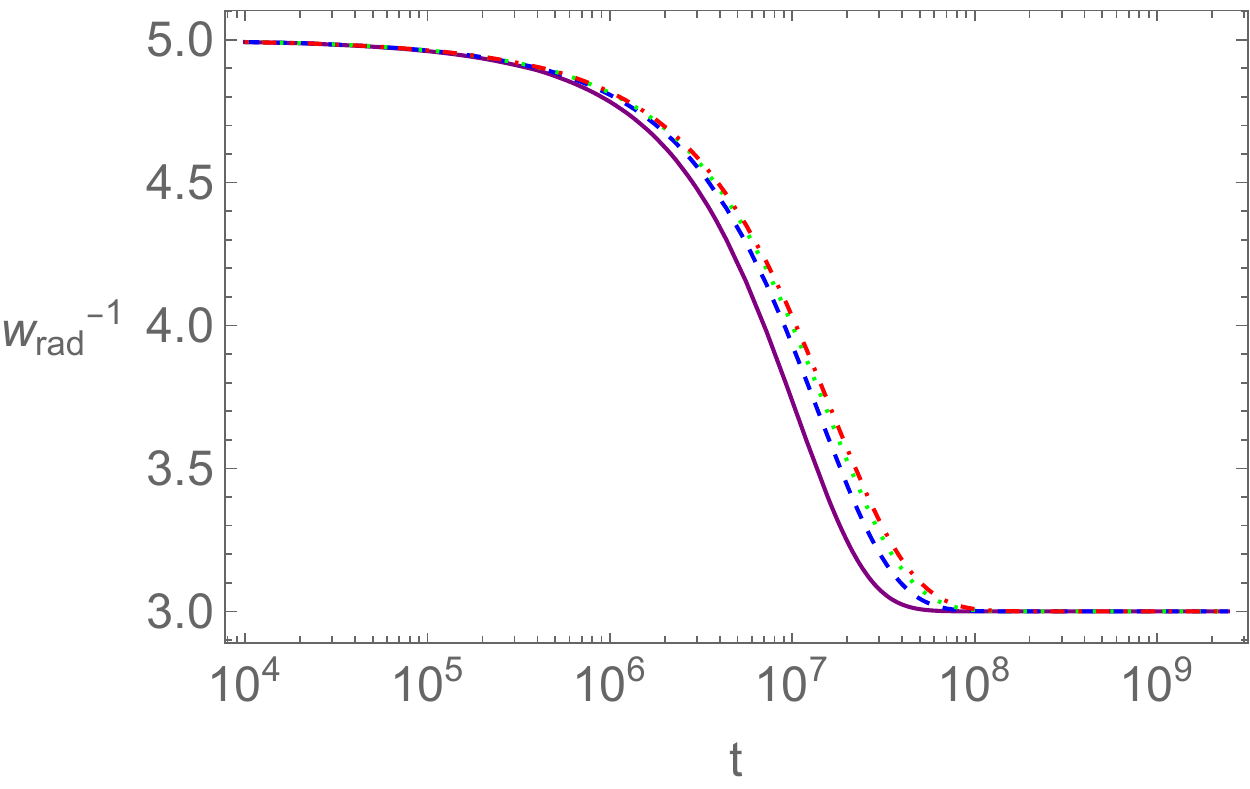}
  \end{center}
\caption{The effective power~$p$ defined in (\ref{def:power}) (left plot) and the reciprocal of $w^{\rm rad}$ defined in (\ref{def:wrad}) (right plot)
as functions of $t$. 
The (purple) solid, (blue) dashed, (green) dotted and (red) dotdashed lines represent the cases of $\beta_{\rm I}=10,12.5,15$ and 20, respectively. 
The constant~$t_0$ in (\ref{def:power}) is chosen as $t_0=-146,-306,-548$ and $-1332$ for those cases. 
The other parameters are chosen as (\ref{ini_values}) with $b_{\rm I}=b_*$ and $\sigma_{\rm I}=\sigma_*$. }
    \label{pwvst}
\end{figure}

Using our approximate expressions, we can see how the time evolution of various quantities 
become the usual 4D ones. 
For example, the ``effective power''~$p$ defined in (\ref{def:power}) is calculated as 
\begin{align}
 p =&\; (t-t_0)\dot{\bar{A}} \simeq \brc{t(x)-t_0}\frac{d\bar{A}}{dx}(x)\brc{\frac{dt}{dx}(x)}^{-1} \nonumber\\
 =&\; \brc{t_{\rm ref}-t_0+\int_{x_{\rm ref}}^x d\tilde{x}\;
 \frac{d\bar{A}}{dx}(\tilde{x})\sqrt{\frac{3}{C_1e^{-3A(\tilde{x})}+\rho^{\rm rad}(\tilde{x})}}}
 \sqrt{\frac{C_1e^{-3A(x)}+\rho^{\rm rad}(x)}{3}}. 
\end{align}
Combining this and (\ref{expr:t}), we obtain $p$ as a function of $t$. 
The constant~$t_0$ is chosen so that $p$ is almost independent of $t$ at earlier times~$t\leq t_{\rm ref}$. 
Similarly we can obtain the equation of state: 
\begin{align}
 w_{\rm rad} \equiv \frac{p_3^{\rm rad}}{\rho^{\rm rad}} 
 \label{def:wrad}
\end{align}
as a function of $t$. 
Notice that its reciprocal~$w_{\rm rad}^{-1}$ measures the effective space dimensions that the radiation feels. 
Fig.~\ref{pwvst} shows the time evolutions of $p$ and $w_{\rm rad}^{-1}$ computed by the above approximations. 
In the case of $\beta_{\rm I}=10$, we can see that $p$ changes from $5/9$, which corresponds to the radiation-dominated 5D universe, 
to 2/3, which corresponds to the oscillating-moduli-dominated 3D universe, as expected. 
For lower initial temperatures, $p$ once decreases and approaches 1/2, which corresponds to the radiation-dominated 3D universe, 
then turns to increase due to the dominance of the moduli oscillation. 
From the right plot of Fig.~\ref{pwvst}, we can see that the effective space dimensions decrease from 5 to 3 
during the period~$10^6<t<10^8$. 
The number of the effective dimensions is reduced to 3 at later times for lower initial temperatures. 

\section{Conditions for radiation dominance}
\label{sec:conditions}
\subsection{Dependence on initial displacements of moduli}
\begin{figure}[t]
  \begin{center}
    \includegraphics[scale=0.65]{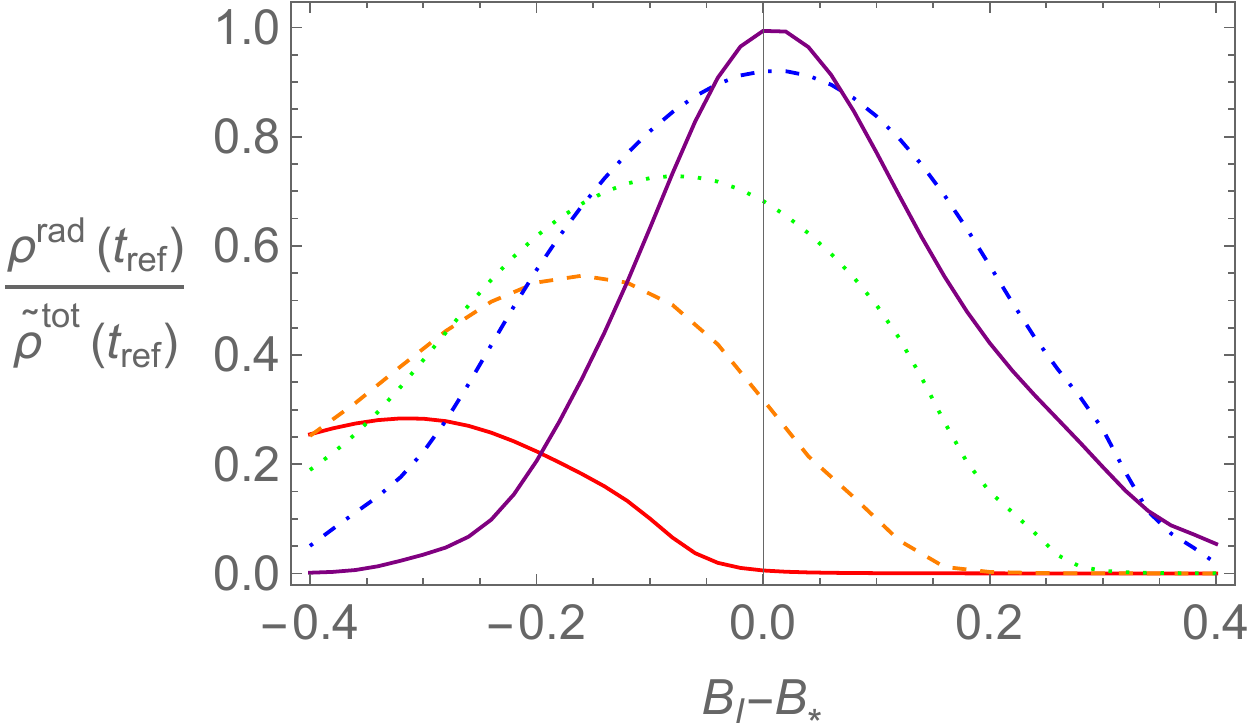} \;\;
    \includegraphics[scale=0.65]{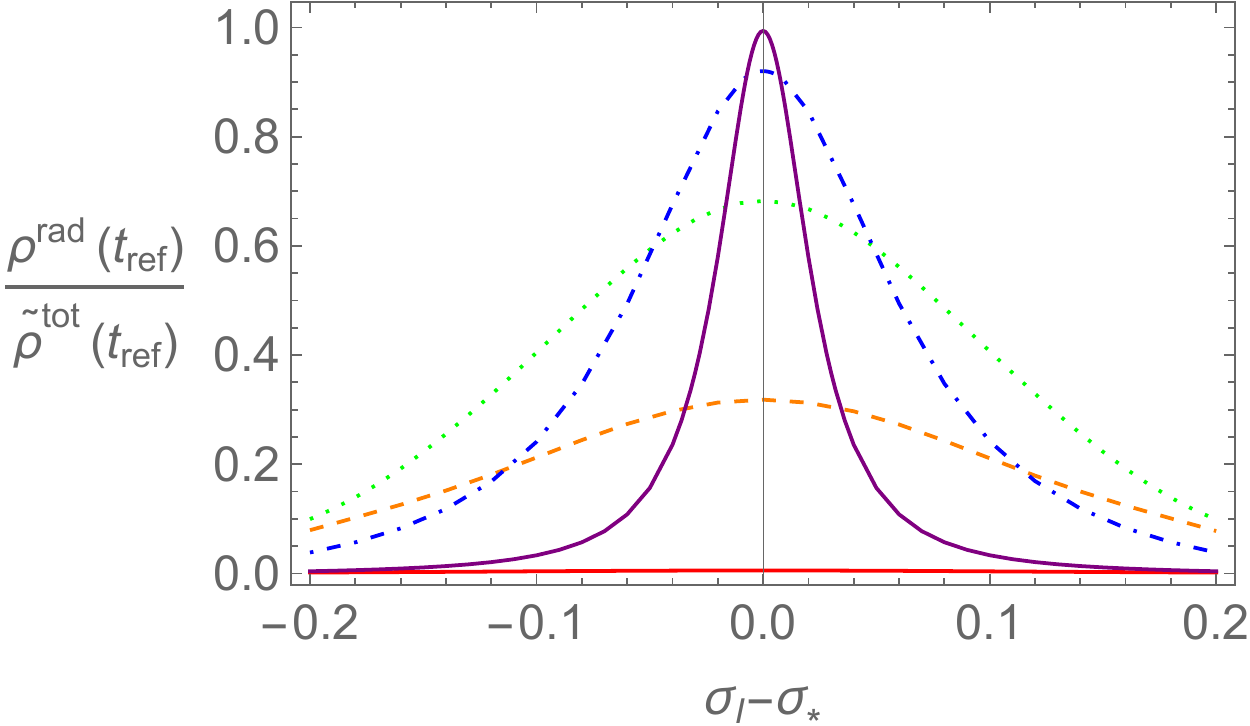} \\
    \vspace{3mm}
    \includegraphics[scale=0.65]{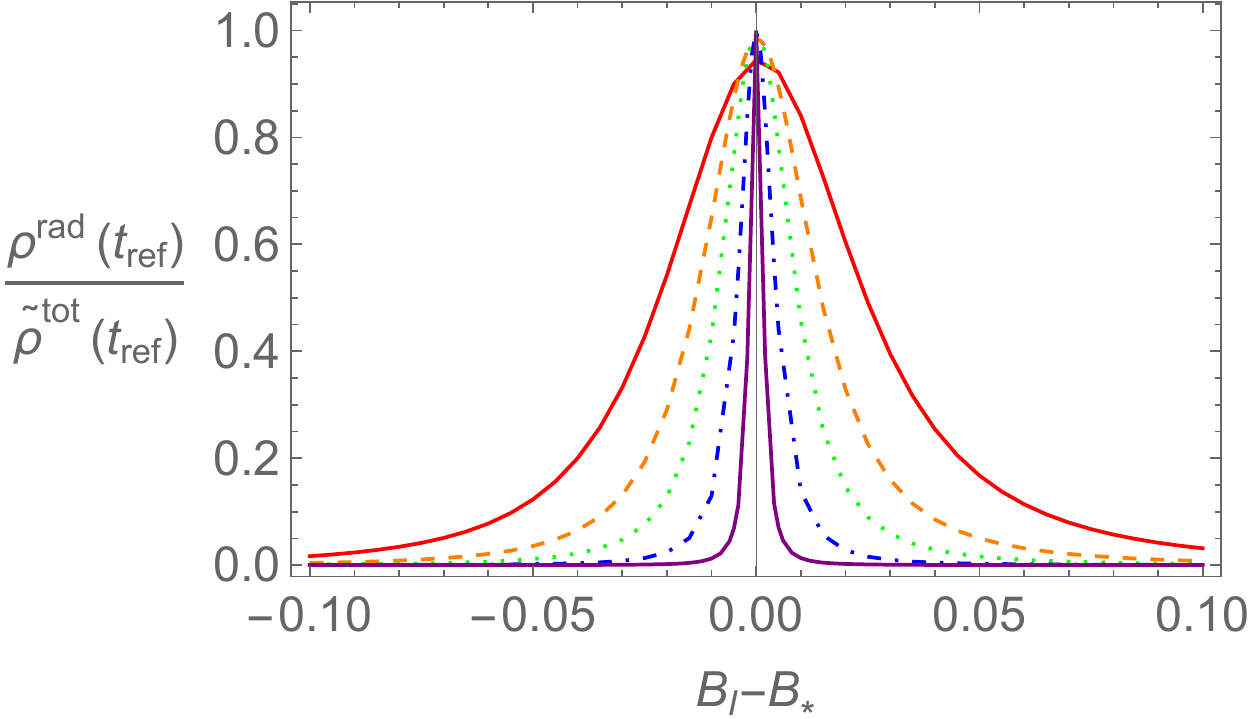} \;\;
    \includegraphics[scale=0.65]{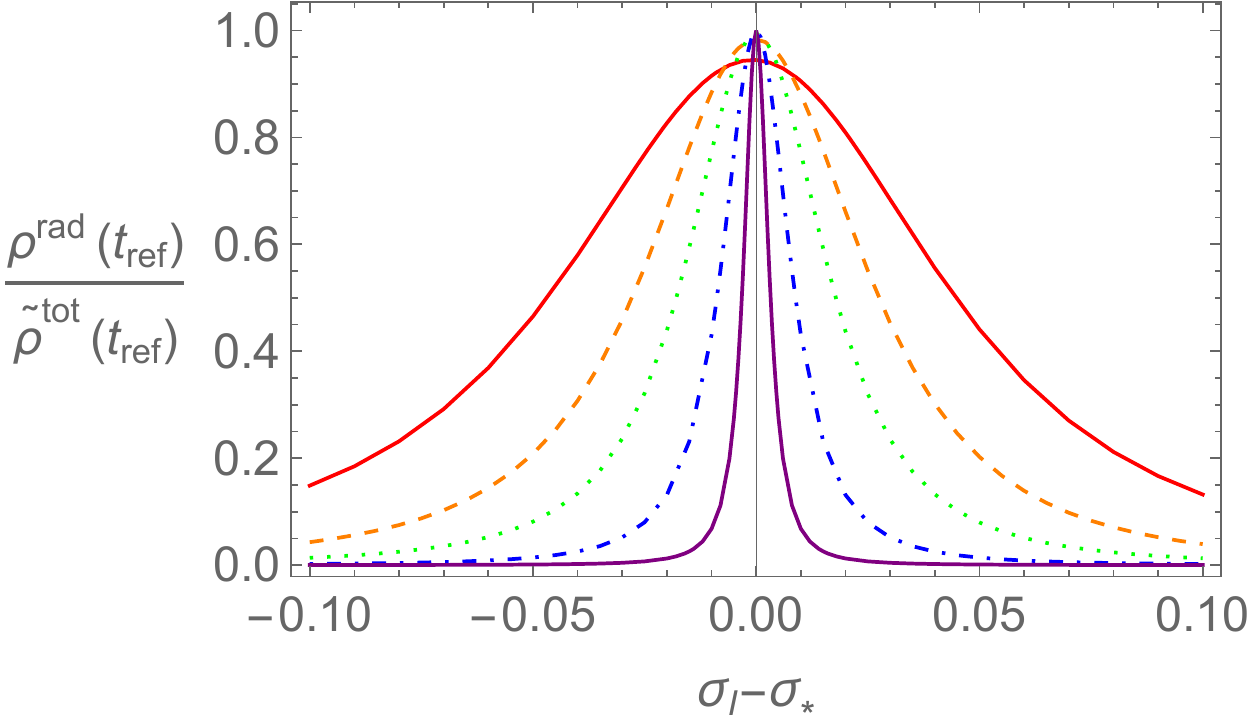}
  \end{center}
\caption{The ratio~$\rho^{\rm rad}/\tilde{\rho}^{\rm tot}$ at $t=t_{\rm ref}$ as functions of $B_{\rm I}-B_*$ with $\sigma_{\rm I}=\sigma_*$ (left plots) 
and $\sigma_{\rm I}-\sigma_*$ with $B_{\rm I}=B_*$ (right plots). 
The upper plots are in the case of $m/m_{\rm KK}^{(1)}=10$ ($\sigma_*=15.9$), 
and the lower plots are in the case of $m/m_{\rm KK}^{(1)}=0.1$ ($\sigma_*=6.6$). 
The red solid, the orange dashed, the green dotted, the blue dotdashed and the purple solid lines correspond 
to $\beta_{\rm I}=10,12.5,15,20$ and 30, respectively. }
    \label{r_ref:B}
\end{figure}

In this subsection, we discuss the dependence of the initial displacement of the moduli~$\Delta B_{\rm I}\equiv B_{\rm I}-B_*$ 
and $\Delta\sigma_{\rm I}\equiv \sigma_{\rm I}-\sigma_*$ on the evolution of the universe. 
As shown in Fig.~\ref{pwvst}, it takes a long time until the moduli oscillation dominates the total energy density of the universe 
when the radiation dominates at $t=t_{\rm ref}$. 
In such a case, if the moduli decay before the oscillation energy dominates, the universe does not experience the moduli-oscillation-dominated era. 
This never happens when the moduli stabilization procedure is described in the 4D EFT 
(i.e., $b_*,\beta\ll m^{-1}$). 
Thus, in this section, we aim to clarify in what case such unusual situations occur. 
Notice that the time evolution of the 3D scale factor~$a$ is determined by the total energy density~$\tilde{\rho}^{\rm tot}$. 
Hence if the ratio~$\rho^{\rm rad}/\tilde{\rho}^{\rm tot}$ is close to one at $t=t_{\rm ref}$, the radiation-dominated era lasts for a long time. 
Fig.~\ref{r_ref:B} shows this ratio at $t=t_{\rm ref}$. 
In the case that $m/m_{\rm KK}^{(1)}=0.1$ (lower plots) and the initial temperature is low, 
the ratio is close to one only when $\Delta B_{\rm I}=\Delta\sigma_{\rm I}=0$, which corresponds to the case that
the moduli have already been stabilized at $t=0$. 
Even small displacements from the stabilized values lead to the dominance of the moduli oscillation. 
This is consistent with the analysis in the 4D EFT. 
In contrast, in the case that $m/m_{\rm KK}^{(1)}=10$ (upper plots), 
the radiation contribution is still non-negligible at $t=t_{\rm ref}$ also for non-zero displacements. 
Note that the moduli stabilization cannot be described in the context of the 4D EFT in this case.  

We should also note that even if the moduli have been stabilized at $t=0$ (i.e., $\Delta B_{\rm I}=\Delta\sigma_{\rm I}=0$), 
the contribution of the moduli oscillation becomes non-negligible at $t=t_{\rm ref}$ 
when $m/m_{\rm KK}^{(1)}\gg 1$ and the initial temperature is high. 
This indicates that the moduli start to oscillate due to the pressure in the extra compact space~$p_2^{\rm rad}$. 
We will discuss this issue in the next subsection.

\subsection{Effects of the pressure~{\boldmath $p_2^{\rm rad}$}}
If no radiation exists, the moduli are safely stabilized for any initial values~$B_{\rm I}$ and $\sigma_{\rm I}$. 
In the presence of the radiation, however, this becomes nontrivial since the pressure in the $S^2$ space~$p_2^{\rm rad}$ 
pushes the moduli from the stabilized values. 
This effect is relevant only at early times because $p_2^{\rm rad}$ is rapidly damped to zero. 
Still, it gives a significant effect on the evolution of the moduli in some cases, 
as we have seen in the upper plots of Fig.~\ref{r_ref:B}. 
In this subsection, we analyze the moduli evolution equations focusing on early times 
and keeping the $p_2^{\rm rad}$ term, which was neglected in Sec.~\ref{Mdl_osc}.

When $\beta\ll b_*$ (i.e., the temperature is much higher than $m_{\rm KK}^{(1)}$), 
the pressure in the compact space~$p_2^{\rm rad}$ is almost independent of the moduli, 
\begin{align}
 \rho^{\rm rad} &\simeq \frac{10g_{\rm dof}}{\pi^3\beta^6}, \;\;\;\;\;
 p_2^{\rm rad} \simeq \frac{2g_{\rm dof}}{\pi^3\beta^6}. 
\end{align}
In this case, the moduli evolution equations become inhomogeneous differential equations. 
Including $p_2^{\rm rad}$, the evolution equation~(\ref{eq:vph12}) is modified as
\begin{align}
 \begin{pmatrix} \ddot{\varphi}_1 \\ \ddot{\varphi}_2 \end{pmatrix} 
 &= -\begin{pmatrix} \lambda_1 & \\ & \lambda_2 \end{pmatrix}\begin{pmatrix} \varphi_1 \\ \varphi_2 \end{pmatrix}
 +2e^{\frac{3}{2}A}p_2^{\rm rad}\begin{pmatrix} \cos\theta \\ -\sin\theta \end{pmatrix}+\cdots, 
 \label{inhomo:vph}
\end{align}
where $\lambda_{1,2}$, $\varphi_{1,2}$ and $\theta$ are defined in (\ref{eigenvalues}), (\ref{def:varphi}) and (\ref{def:theta}). 
We have neglected terms involving $\ddot{A}$. 
From (\ref{rel:dotbt-dota}) and the fact that $v_\beta(x)\simeq 3/5$ for $x\ll 1$, 
we can express the inverse temperature as $\beta\simeq \beta_{\rm I}a^{3/5}$, and thus obtain
\begin{align}
 \rho^{\rm rad} &\simeq \frac{10g_{\rm dof}}{\pi^3\beta_{\rm I}^6}e^{-\frac{18}{5}\bar{A}} 
 \equiv C_2e^{-\frac{18}{5}\bar{A}}, \nonumber\\
 p_2^{\rm rad} &\simeq \frac{2g_{\rm dof}}{\pi^3\beta_{\rm I}^6}e^{-\frac{18}{5}\bar{A}} = \frac{C_2}{5}e^{-\frac{18}{5}\bar{A}}. 
\end{align} 

If we assume that $\dot{\bar{A}}(t)\leq\dot{\bar{A}}(0)\ll \lambda_1$, 
(\ref{expr:cstrt:1}) can be approximated as (\ref{3barA^2}). 
Thus the energy density of the moduli oscillation is estimated as 
\begin{align}
 \rho^{\rm osc} &\equiv \tilde{\rho}^{\rm tot}-\rho^{\rm rad} 
 = \frac{e^{-3\bar{A}}}{2}\brkt{\dot{\varphi}_1^2+\dot{\varphi}_2^2+\lambda_1\varphi_1^2+\lambda_2\varphi_2^2}. 
 \label{def:rho^osc}
\end{align}
Here we assume that 
\begin{align}
 C_{\rm 1I} &\equiv \frac{1}{2}\brkt{\dot{\varphi}_1^2+\dot{\varphi}_2^2+\lambda_1\varphi_1^2+\lambda_2\varphi_2^2} 
\end{align}
is almost independent of the time~$t$. 
Then, similarly to (\ref{dotbarA}), we can express $\dot{\bar{A}}$ as
\begin{align}
 \dot{\bar{A}} &\simeq \sqrt{\frac{1}{3}\brkt{C_{\rm 1I}e^{-3\bar{A}}+C_2e^{-\frac{18}{5}\bar{A}}}}, 
\end{align}
which leads to 
\begin{align}
 t &\simeq \int_1^{\bar{A}} dz\;\sqrt{\frac{3}{C_{\rm 1I}e^{-3z}+C_2e^{-\frac{18}{5}z}}} 
 \simeq \begin{cases} \displaystyle \frac{2}{\sqrt{3C_{\rm 1I}}}\brkt{e^{\frac{3}{2}\bar{A}}-1} & \mbox{for $C_{\rm 1I} \gg C_2$} \\
 \displaystyle \frac{5}{3\sqrt{3C_2}}\brkt{e^{\frac{9}{5}\bar{A}}-1} & \mbox{for $C_{\rm 1I}e^{\frac{3}{5}\bar{A}} \ll C_2$} \end{cases}. 
\end{align}
By taking the inverse function, the 3D scale factor is expressed as 
\begin{align}
 a(t) &= e^{A(t)} \simeq \begin{cases} \displaystyle \brkt{1+\frac{\sqrt{3C_{\rm 1I}}}{2}t}^{2/3} & \mbox{for $C_{\rm 1I}\gg C_2$} \\
 \displaystyle \brkt{1+\frac{3\sqrt{3C_2}}{5}t}^{5/9} & \mbox{for $C_{\rm 1I}e^{\frac{3}{5}\bar{A}} \ll C_2$} \end{cases}, 
 \label{ini:a-t:ap}
\end{align}
Therefore, we find the time dependence of the inhomogeneous term in (\ref{inhomo:vph}) as
\begin{align}
 2e^{\frac{3}{2}A}p_2^{\rm rad} &\simeq \frac{4g_{\rm dof}}{\pi^3\beta_{\rm I}^6}e^{-\frac{21}{10}\bar{A}(t)} 
 \simeq \begin{cases} \displaystyle \frac{2C_2}{5}\brkt{1+\frac{\sqrt{3C_{\rm 1I}}}{2}t}^{-\frac{7}{5}} & \mbox{for $C_{\rm 1I} \gg C_2$} \\
 \displaystyle \frac{2C_2}{5}\brkt{1+\frac{3\sqrt{3C_2}}{5}t}^{-\frac{7}{6}} & \mbox{for $C_{\rm 1I}e^{\frac{3}{5}\bar{A}} \ll C_2$} \end{cases}. 
\end{align}

To illustrate the situation, we roughly approximate this as
\begin{align}
 2e^{\frac{3}{2}A}p_2^{\rm rad} &\sim \frac{2C_2}{5}\brkt{1+\alpha t}^{-1}, 
\end{align}
where $\alpha\sim \sqrt{C_{\rm 1I}}$ for $C_{\rm 1I}\gg C_2$, 
and $\alpha\sim\sqrt{C_2}$ for $C_{\rm 1I}\leq C_{\rm 1I}e^{3\bar{A}/5}\ll C_2$.  
Then (\ref{inhomo:vph}) can be solved as
\begin{align}
 \varphi_1(t) &= \varphi_{\rm 1I}\brc{\cos\brkt{\sqrt{\lambda_1}t}+\frac{3H_{\rm I}}{2\sqrt{\lambda_1}}\sin\brkt{\sqrt{\lambda_1}t}}
 -\frac{2C_2\cos\theta}{5\alpha\sqrt{\lambda_1}}{\cal G}\brkt{\sqrt{\lambda_1}t;\frac{\sqrt{\lambda_1}}{\alpha}},
 \nonumber\\
 \varphi_2(t) &= \varphi_{\rm 2I}\brc{\cos\brkt{\sqrt{\lambda_2}t}+\frac{3H_{\rm I}}{2\sqrt{\lambda_2}}\sin\brkt{\sqrt{\lambda_2}t}}
 +\frac{2C_2\sin\theta}{5\alpha\sqrt{\lambda_2}}{\cal G}\brkt{\sqrt{\lambda_2}t;\frac{\sqrt{\lambda_2}}{\alpha}}, 
 \label{t-evolv:varphi}
\end{align}
where
\begin{align}
 {\cal G}(z;c) &\equiv \brc{{\rm Si}(z+c)-{\rm Si}(c)}\cos(z+c)-\brc{{\rm Ci}(z+c)-{\rm Ci}(c)}\sin(z+c). 
\end{align}
The functions~${\rm Si}(z)$ and ${\rm Ci}(z)$ are trigonometric integrals defined by
\begin{align}
 {\rm Si}(z) &\equiv \int_0^zdw\;\frac{\sin w}{w}, \;\;\;\;\;
 {\rm Ci}(z) \equiv -\int_z^\infty dw\;\frac{\cos w}{w}. 
\end{align}
We have used the initial conditions, 
\begin{align}
 \varphi_1(0) &= \varphi_{\rm 1I} 
 \equiv 2\Delta B_{\rm I}\cos\theta+\Delta\sigma_{\rm I}\sin\theta, \;\;\;\;\;
 \dot{\varphi}_1(0) = \frac{3}{2}\dot{A}(0)\varphi_{\rm 1I}, \nonumber\\
 \varphi_2(0) &= \varphi_{\rm 2I} 
 \equiv -2\Delta B_{\rm I}\sin\theta+\Delta\sigma_{\rm I}\cos\theta, \;\;\;\;\;
 \dot{\varphi}_2(0) = \frac{3}{2}\dot{A}(0)\varphi_{\rm 2I}, \nonumber\\
 \dot{A}(0) &= \sqrt{\frac{1}{3}\brc{2e^{-\sigma_{\rm I}}\brkt{1-e^{\Delta\sigma_{\rm I}-\Delta B_{\rm I}}}^2
 +\frac{m^2}{2}(\Delta \sigma_{\rm I})^2+C_2}} 
 \equiv H_{\rm I}, 
\end{align}
where $\Delta B_{\rm I}\equiv B_{\rm I}-B_*$ and $\Delta\sigma_{\rm I}\equiv \sigma_{\rm I}-\sigma_*$. 

In order to see the impact of $p_2^{\rm rad}$ on the spacetime evolution at early times, 
we focus on the case of $\Delta B_{\rm I}=\Delta\sigma_{\rm I}=0$. 
Namely, the moduli have already been stabilized at the beginning of the radiation-dominated era. 
Then the energy density of the moduli oscillation~(\ref{def:rho^osc}) is estimated as 
\begin{align}
 \rho^{\rm osc} 
 &= \frac{2C_2^2}{25\alpha^2}e^{-3\bar{A}(t)}\brc{\cos^2\theta{\cal H}\brkt{\sqrt{\lambda_1}t;\frac{\sqrt{\lambda_1}}{\alpha}}
 +\sin^2\theta{\cal H}\brkt{\sqrt{\lambda_2} t;\frac{\sqrt{\lambda_2}}{\alpha}}}, 
 \label{rho^osc}
\end{align}
where 
\begin{align}
 {\cal H}(z;c) &\equiv {\cal G}^2(z;c)+\brc{{\cal G}'(z;c)}^2 \nonumber\\
 &= \brc{{\rm Ci}\,(z+c)-{\rm Ci}\,(c)}^2+\brc{{\rm Si}\,(z+c)-{\rm Si}\,(c)}^2. 
\end{align}
\begin{figure}[t]
  \begin{center}
    \includegraphics[scale=0.65]{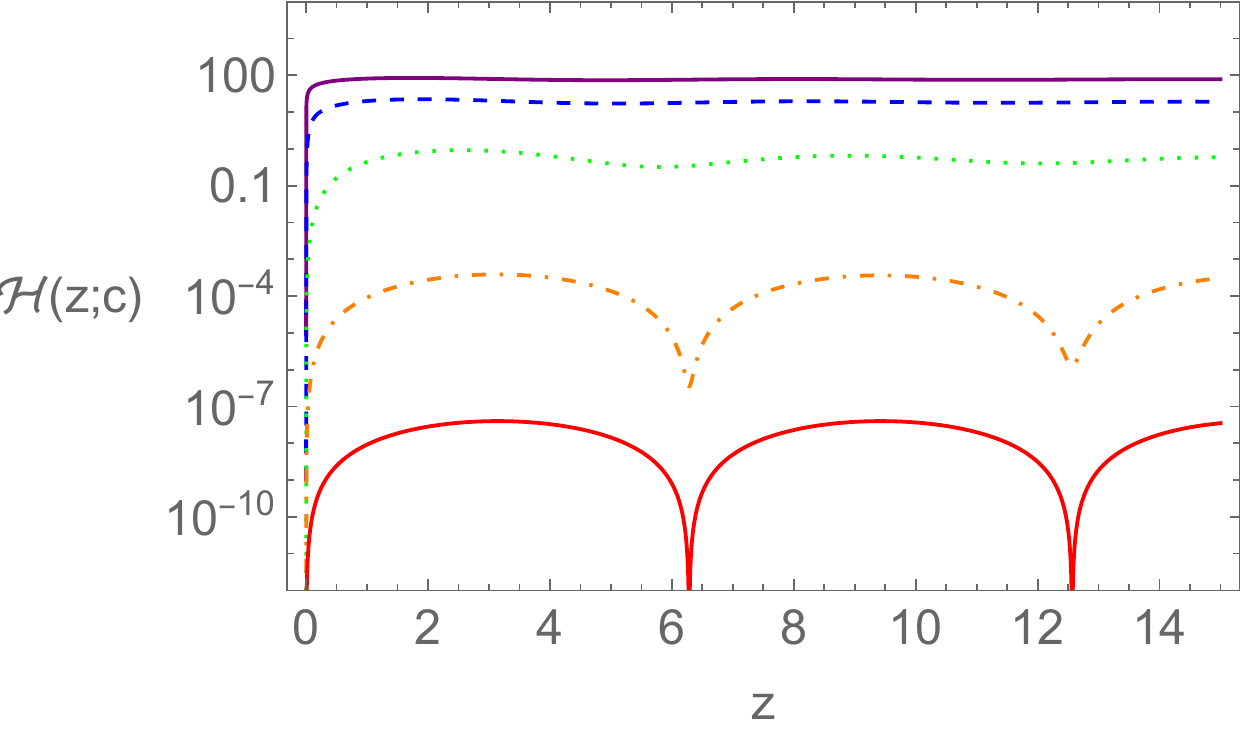} \;\;
    \includegraphics[scale=0.65]{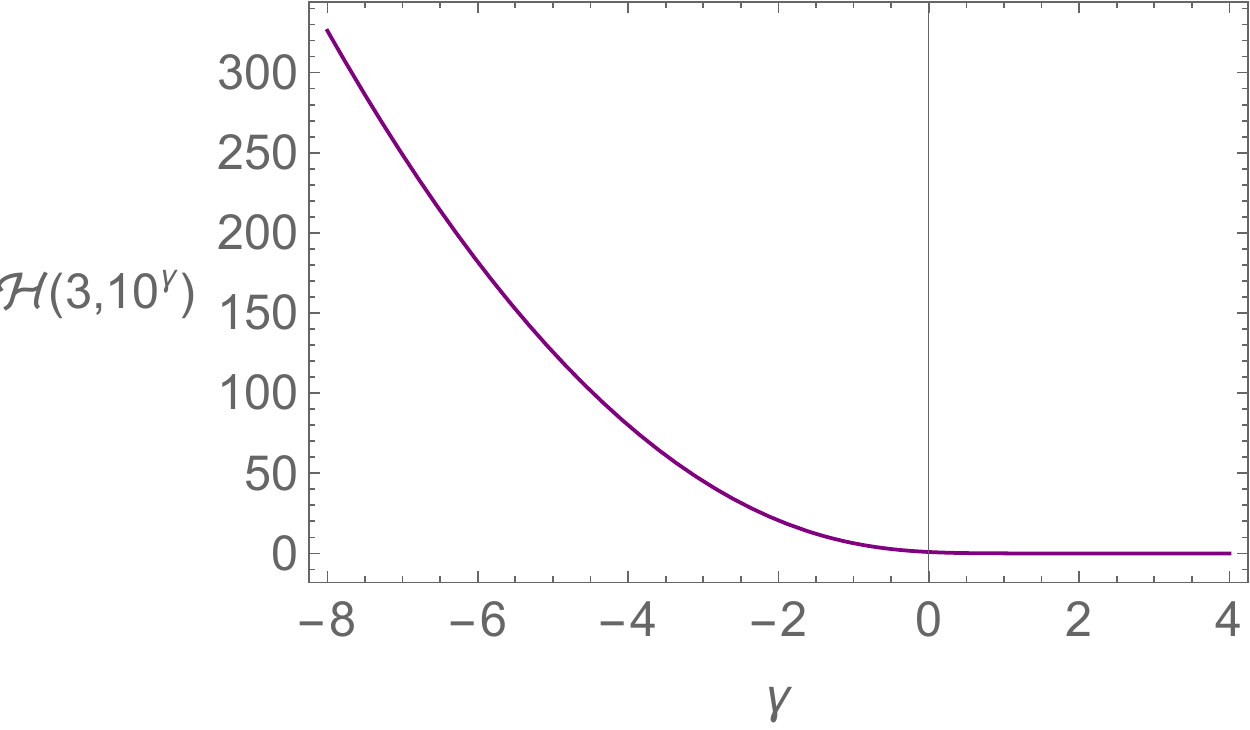} \\
  \end{center}
\caption{The profiles of ${\cal H}(z;c)$. In the left plot, the parameter~$c$ is chosen as $c=10^{-4},10^{-2},1,10^2,10^4$ 
from top to bottom. The right plot shows the first peak value of ${\cal H}(z;c)$ as a function of $\gamma\equiv \ln_{10}c$. 
}
    \label{cHs}
\end{figure}
Fig.~\ref{cHs} shows the profiles of the function~${\cal H}(z;c)$. 
As we can see from the left plot, ${\cal H}(z;c)$ is almost independent of $z$ when $c\ll 1$. 
In that case, ${\cal H}(z;c)\simeq \ln^2c$ (see the right plot).  
When $c>1$ ($z>0$), ${\cal H}(z;c)$ is an oscillating function around $1/c^2$, 
whose first peak is around $z=3$ and ${\cal H}(3;c)\simeq 4/c^2$. 

Here let us assume that 
\begin{align}
 C_{\rm 1I}\ll C_2. 
 \label{assump1}
\end{align}
Then, since $\alpha\sim \sqrt{C_2}$, (\ref{rho^osc}) becomes
\begin{align}
 \rho^{\rm osc}(t) &\sim \frac{2C_2}{25}e^{-3\bar{A}(t)}\brc{\cos^2\theta{\cal H}\brkt{\sqrt{\lambda_1}t;\sqrt{\frac{\lambda_1}{C_2}}}
 +\sin^2\theta{\cal H}\brkt{\sqrt{\lambda_2}t;\sqrt{\frac{\lambda_2}{C_2}}}}. 
\end{align}
Thus if 
\begin{align}
 \rho^{\rm osc}(t_{\rm ref}) &\sim \frac{2C_2e^{-3\bar{A}(t_{\rm ref})}}{25}
 \brc{\cos^2\theta{\cal H}\brkt{\sqrt{\lambda_1}t_{\rm ref};\sqrt{\frac{\lambda_1}{C_2}}}
 +\sin^2\theta{\cal H}\brkt{\sqrt{\lambda_2}t_{\rm ref};\sqrt{\frac{\lambda_2}{C_2}}}} \nonumber\\
 &\ll \rho^{\rm rad}(t_{\rm ref}) \simeq C_2e^{-\frac{18}{5}\bar{A}(t_{\rm ref})}  
\end{align}
holds, the assumption~(\ref{assump1}) is justified. 
This condition is rewritten as
\begin{align}
 {\cal R} \ll 1, 
 \label{eq:crit}
\end{align}
where~\footnote{ 
We have used the approximate expression in (\ref{ini:a-t:ap}), 
and $3\sqrt{3}/5\simeq 1$. 
}
\begin{align}
 {\cal R} &\equiv \frac{2}{25}\brkt{1+\sqrt{C_2}t_{\rm ref}}^{1/3} 
 \brc{\cos^2\theta{\cal H}\brkt{\sqrt{\lambda_1}t_{\rm ref};\sqrt{\frac{\lambda_1}{C_2}}}
 +\sin^2\theta{\cal H}\brkt{\sqrt{\lambda_2}t_{\rm ref};\sqrt{\frac{\lambda_2}{C_2}}}}. 
 \label{criterion}
\end{align}
In this case, the radiation energy is dominated at $t=t_{\rm ref}$. 
If (\ref{eq:crit}) does not hold, the contribution of the moduli oscillation to the total energy density cannot be negligible, 
and it will dominate after $t=t_{\rm ref}$.  
Note that $\lambda_1$, $\lambda_2$ and $\theta$ are functions of $m$ and $\sigma_*$ (or $m_{\rm KK}^{(1)}$), 
and $C_2$ is a function of $\beta_{\rm I}$. 

\begin{figure}[t]
  \begin{center}
    \includegraphics[scale=0.6]{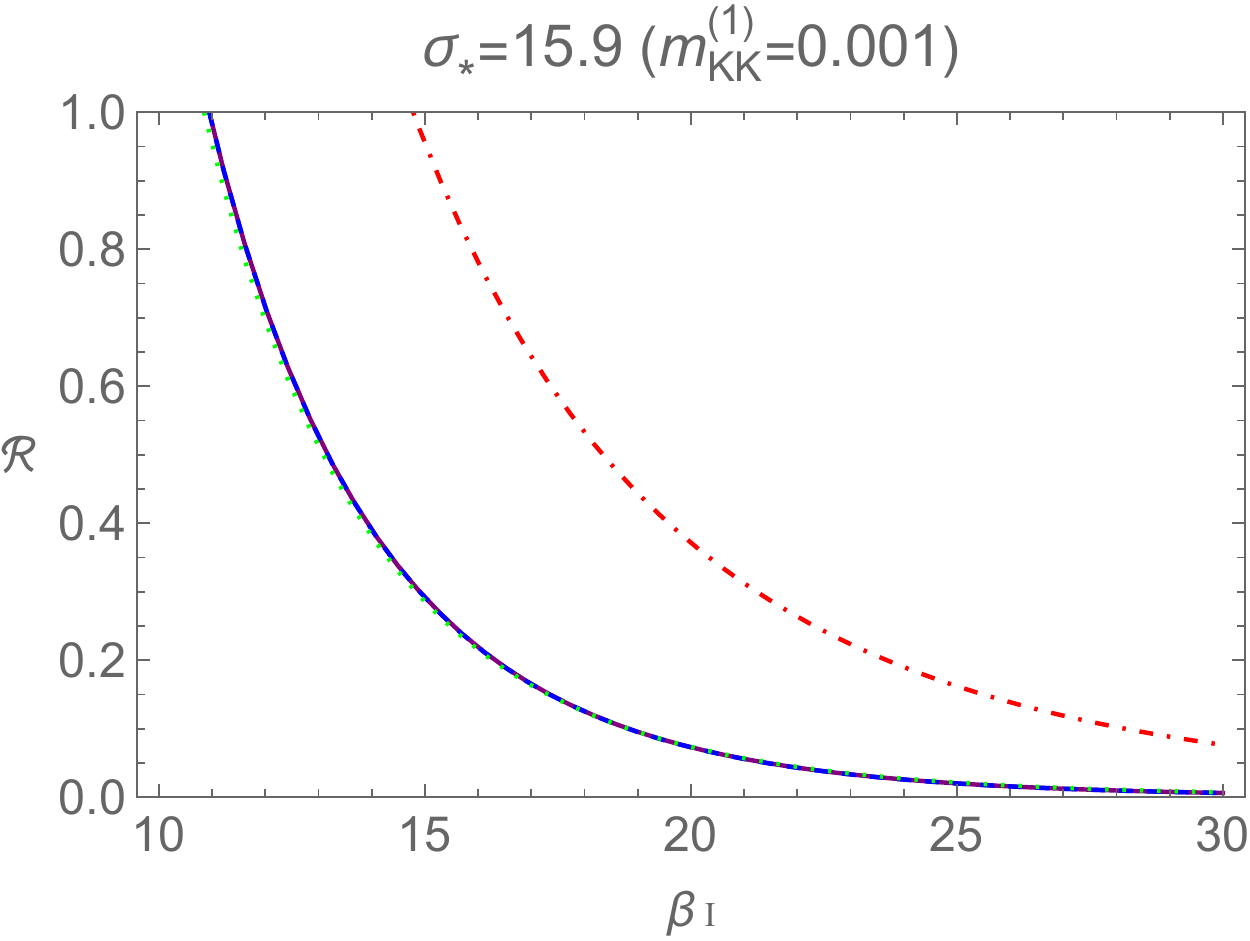}  \;\;
    \includegraphics[scale=0.6]{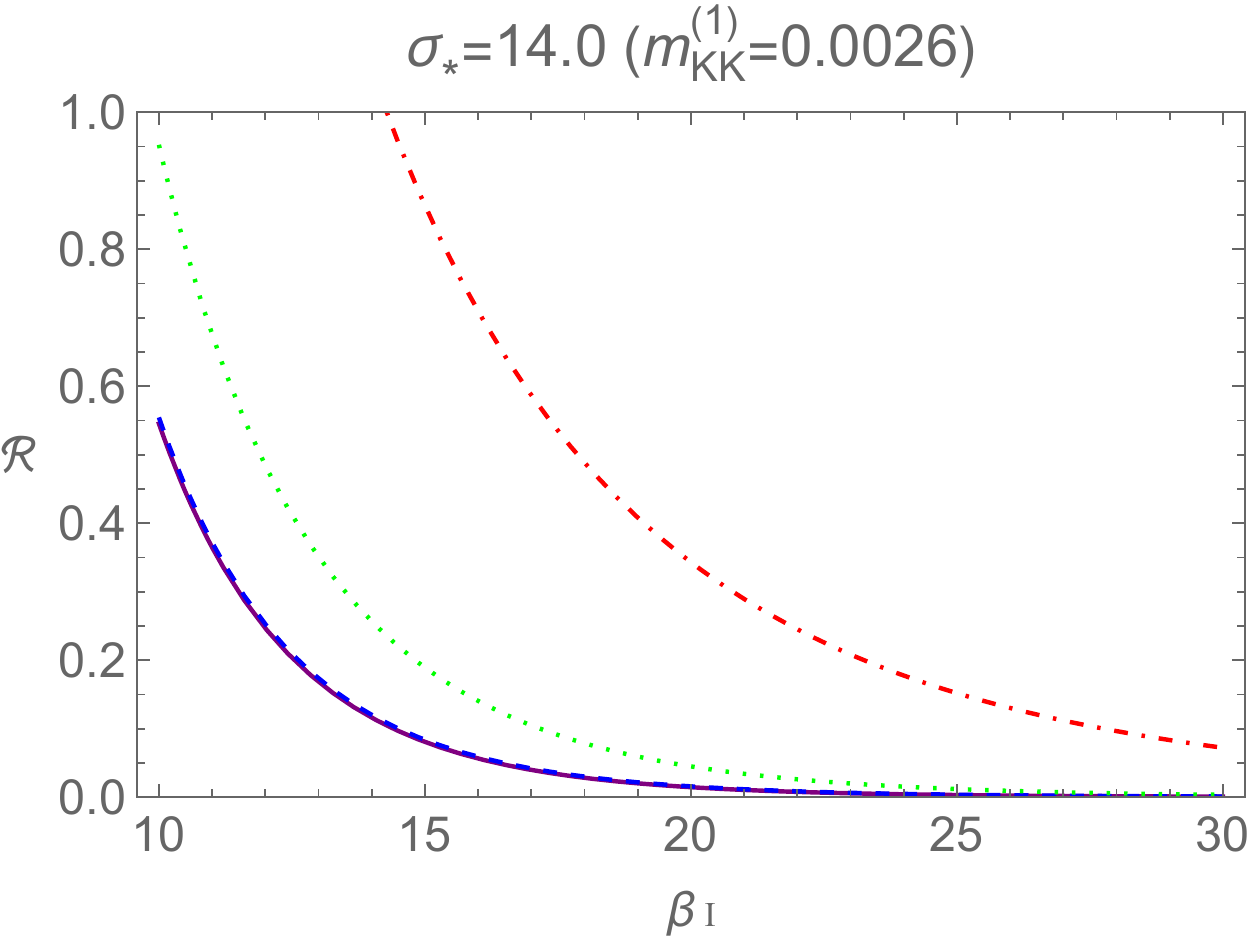} \\
    \includegraphics[scale=0.6]{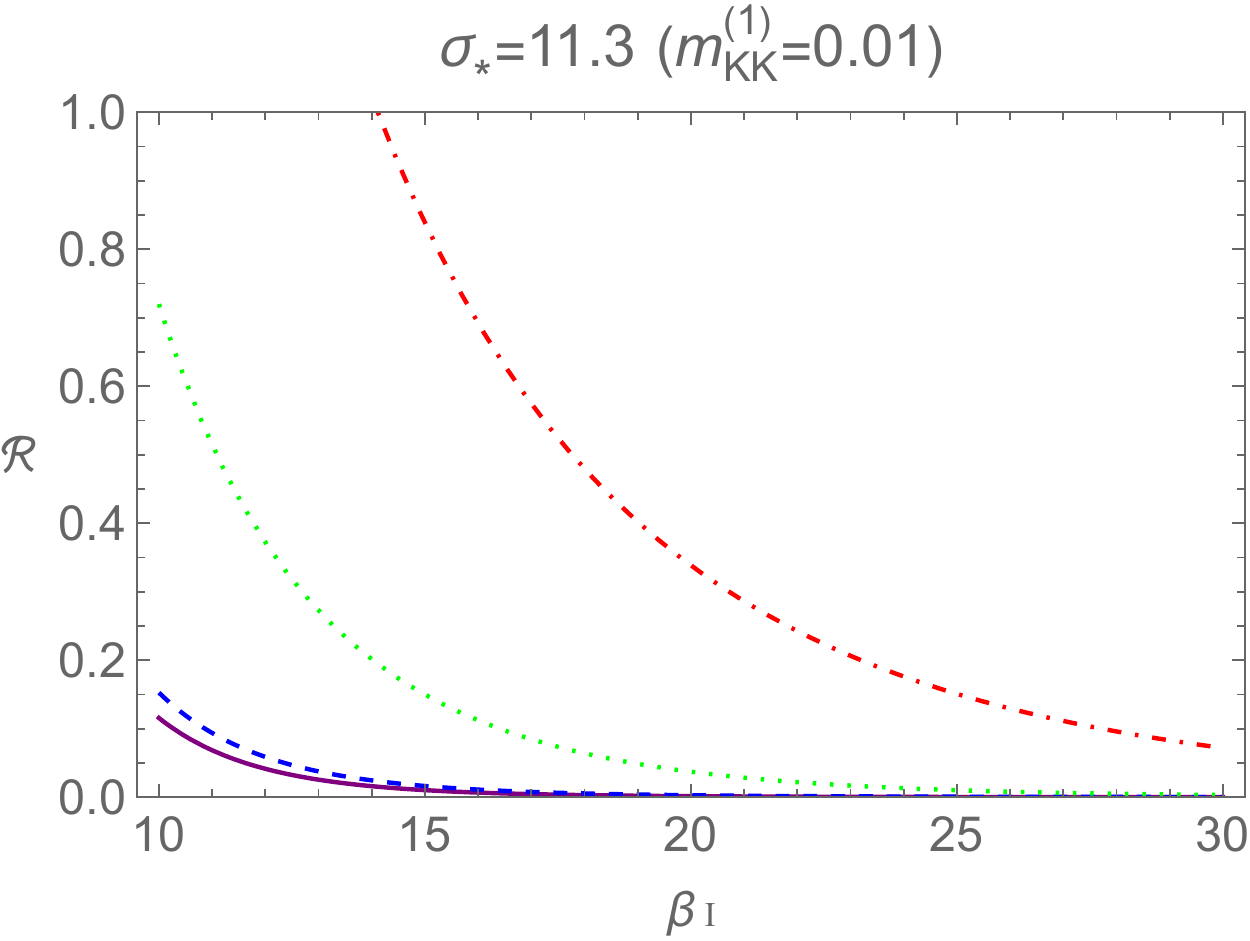} \;\;
    \includegraphics[scale=0.6]{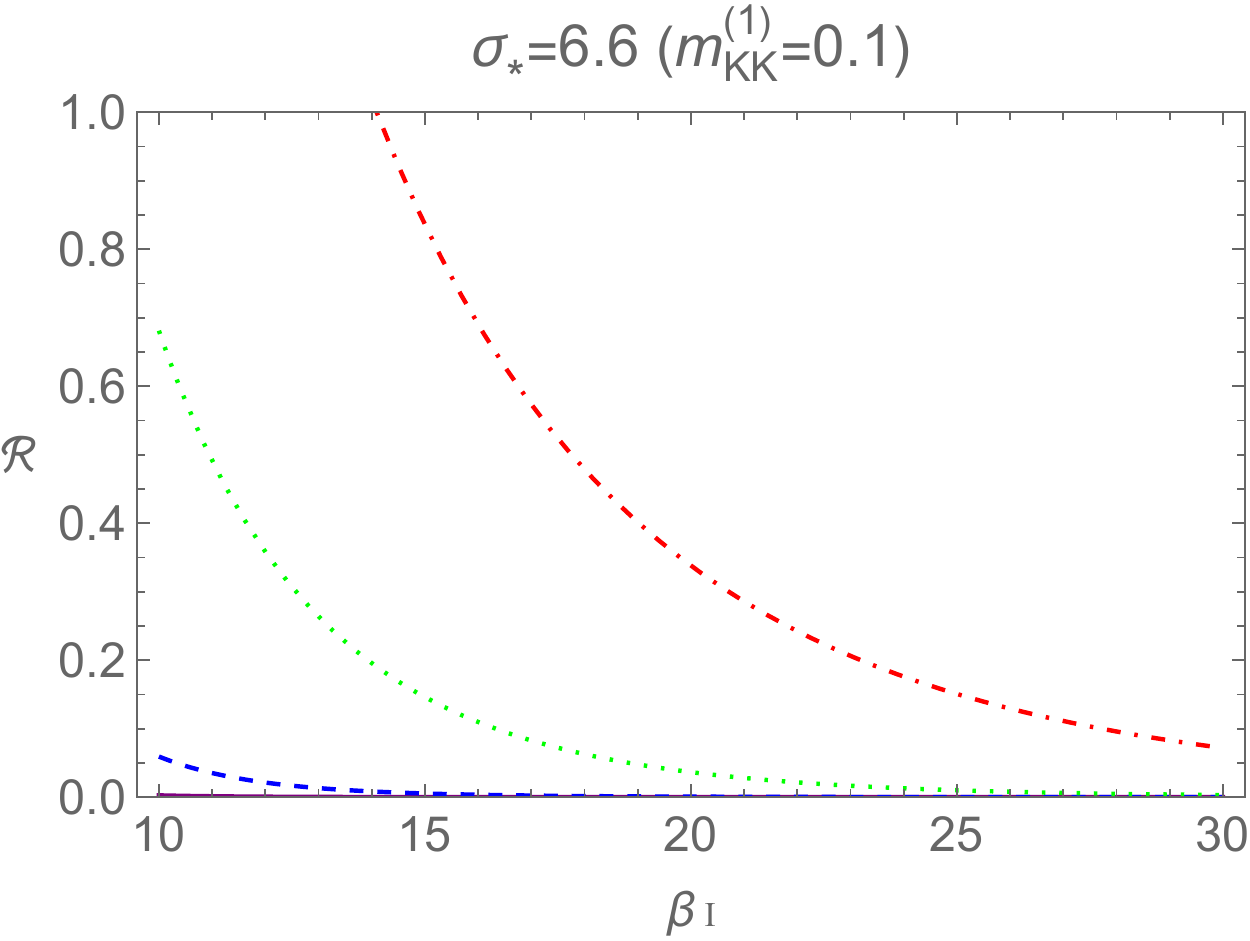}
  \end{center}
\caption{${\cal R}$ as a function of $\beta_{\rm I}$. 
The (purple) solid, (blue) dashed, (green) dotted, and (red) dotdashed lines correspond to $m=0.1$, 0.01, 0.001 and 0.0001, respectively. 
}
    \label{cRs}
\end{figure}
Fig.~\ref{cRs} shows ${\cal R}$ as a function of the initial (inverse) temperature~$\beta_{\rm I}$ 
for various values of $m$ and $\sigma_*$ (or $m_{\rm KK}^{(1)}$). 
The other parameters are chosen as $g_{\rm dof}=100$ and $t_{\rm ref}=10000$. 
We can see that the (blue) dashed lines in the upper left and the lower right plots are consistent with 
the result shown in Fig.~\ref{r_ref:B}. 
In the case of $m=0.01$ and $m_{\rm KK}^{(1)}=0.001$, the contribution of the moduli oscillation becomes non-negligible 
for $\beta\sim 10$ even when the initial moduli values are set at the stabilized values. 
In contrast, in the case of $m=0.01$ and $m_{\rm KK}^{(1)}=0.1$, the radiation contribution always dominates. 

When $m\gg m_{\rm KK}^{(1)}$, $\lambda_1\simeq m_{\rm KK}^{(1)2}$, $\lambda_2\simeq m^2$ and $\tan\theta\ll 1$ 
(see (\ref{eigenvalues}) and (\ref{def:theta})). 
In this case, (\ref{criterion}) reduces to 
\begin{align}
 {\cal R} &\simeq \frac{2}{25}\brkt{1+\sqrt{\frac{10g_{\rm dof}}{\pi^3}}\frac{t_{\rm ref}}{\beta_{\rm I}^3}}^{1/3}
 {\cal H}\brkt{m_{\rm KK}^{(1)}t_{\rm ref};\sqrt{\frac{\pi^3}{10g_{\rm dof}}}m_{\rm KK}^{(1)}\beta_{\rm I}^3}, 
\end{align}
which is almost independent of $\lambda_2\simeq m^2$. 
Recalling the behavior of ${\cal H}(z;c)$ (see the right plot in Fig.~\ref{cHs}), ${\cal R}$ is estimated as 
\begin{align}
 {\cal R} &\sim \begin{cases} \displaystyle \ln^2\brkt{m_{\rm KK}^{(1)}\beta_{\rm I}^3} & \brkt{m_{\rm KK}^{(1)}\beta_{\rm I}^3 \ll 1} \\
 \displaystyle \frac{1}{m_{\rm KK}^{(1)2}\beta_{\rm I}^6} \ll 1 & \brkt{m_{\rm KK}^{(1)}\beta_{\rm I}^3 \gg 1} \end{cases}. 
\end{align}
We have assumed that the factor in front of ${\cal H}$ is ${\cal O}(1)$. 
Therefore, the moduli oscillation dominates the total energy density when $m_{\rm KK}^{(1)}\beta_{\rm I}^3\ll 1$, 
while the its contribution is negligible when $m_{\rm KK}^{(1)}\beta_{\rm I}^3\gg 1$. 
We can see these properties from the upper plots in Fig.~\ref{cRs}.

\section{Summary}
\label{sec:con}
We analyze spacetime evolution during the moduli oscillation in a 6D model compactified on $S^2$ in the presence of the radiation. 
In our previous work~\cite{Otsuka:2022rpx}, we studied the model by numerically solving the field equations, 
and found that the radiation contribution to the total energy density remains non-negligible for a long time 
in a situation that the moduli stabilization dynamics cannot be described in the context of 4D EFT. 
This is in contrast to the result obtained by the conventional 4D EFT approach. 
However, such a numerical approach is available for only a limited range of the time. 
In fact, it is practically difficult to pursue the evolution until the spacetime behaves like 4D. 
In the current work, we develop a procedure to compute it for large $t$, and see the transition from 6D to 4D explicitly. 

The evolution of the 3D scale factor is characterized by the power~$p$ defined in (\ref{def:power}). 
When the initial temperature is high enough, $p$ monotonically increases up to 2/3, 
which corresponds to the moduli-oscillation-dominated 4D universe. 
For lower initial temperatures, it first decreases and approaches 1/2, which corresponds to the radiation-dominated 4D universe, 
and then turns to increase. 
This indicates that the radiation feels the decrease of the effective space dimensions before the moduli oscillation dominates 
the total energy density. 
In such a case, the universe never experiences the moduli-oscillation-dominated era 
if the moduli decay before the dominance of the moduli oscillation. 

We also study the conditions that the radiation remains to give a dominant contribution to the total energy density for a long time. 
We found that such a situation occurs when the moduli stabilization dynamics cannot be described in the context of 4D EFT 
(i.e., $b_*,\beta\gg m^{-1}$). 

Another important result we obtained is that even if the moduli have already been stabilized at the beginning of the radiation-dominated era, 
the pressure in the extra compact space~$p_2^{\rm rad}$ pushes the stabilized moduli and they start to oscillate again in some cases. 
This occurs when the KK mass scale and the initial temperature are higher than the mass scale of the moduli stabilization potential. 
When the moduli stabilization is described in the context of 4D EFT, 
this never happens because $p_2^{\rm rad}$ is negligible in such a case. 

In this work, we have neglected the decay of the moduli, which is essential to the study of realistic scenarios.  
In the subsequent papers, we will take it into account and investigate a full thermal history of the
universe, incorporating the Standard Model sector localized on a codimension-two
brane~\cite{Aghababaie:2003wz}-\cite{Lee:2005az}.

\subsection*{Acknowledgements}

  H. O. was supported in part by JSPS KAKENHI Grant Numbers JP20K14477 and the Education and Research Program for Mathematical and Data Science from the Kyushu University.

\appendix

\section{Thermodynamic quantities}
\label{TDquantities}
The dispersion relation of a 6D relativistic or massless particle is 
\begin{align}
k^Mk_M &= -k_0^2+\frac{1}{a^2}\vec{k}^2+\frac{1}{b^2}k_\theta^2+\frac{1}{b^2\sin^2\theta}k_\phi^2 = 0. 
\end{align}
Thus the energy of the particle with the 3D momentum~$\vec{k}=(k_1,k_2,k_3)$ and the angular momentum~$l$ on $S^2$ is given by
\begin{align}
 {\cal E}_{k,l} &= k_0 = \sqrt{\frac{k^2}{a^2}+\frac{l(l+1)}{b^2}}, 
\end{align}
where $k\equiv \sqrt{\vec{k}^2}$. 
Since each one-particle state is specified by $\vec{k}$, $l$ and the `magnetic quantum number'~$m=-l,\cdots,l$, 
we have $(2l+1)$ degenerate energy eigenstates for each $\vec{k}$ and $l$. 
Hence the grand potential is expressed as 
\begin{align}
 J(\beta,\mu,{\cal V}_3,{\cal V}_2) &= \pm\sum_{l=0}^\infty\frac{g_{\rm dof}(2l+1)}{2\pi^2\beta}
 \int_0^\infty dk\;k^2\ln\brkt{1\mp e^{-\beta({\cal E}_{k,l}-\mu)}} \nonumber\\
 &= \mp\frac{g_{\rm dof}{\cal V}_3}{\pi^2\beta^4}{\rm Li}_4(\pm e^{\beta\mu})
 \pm\sum_{l=1}^\infty\frac{g_{\rm dof}(2l+1){\cal V}_3}{2\pi^2\beta^4}
 \int_0^\infty dq\;q^2\ln\brkt{1\mp e^{-\sqrt{q^2+c_l^2}+\beta\mu}}, 
 \label{def:J}
\end{align}
where $g_{\rm dof}$ denotes the degrees of freedom for the 6D relativistic particles, $\beta$ is the inverse temperature, 
$\mu$ is the chemical potential, and ${\cal V}_3\equiv a^3$ and ${\cal V}_2\equiv4\pi b^2$ are the comoving volume for the 3D space 
and the physical volume of $S^2$, respectively. 
The upper (lower) signs correspond to the case of bosons (fermions). 
At the second equality, we have rescaled the integration variable and the KK masses as 
\begin{align}
 q &\equiv \frac{\beta}{a}k, \;\;\;\;\;
 c_l \equiv \beta\sqrt{\frac{4\pi l(l+1)}{{\cal V}_2}} = \frac{\beta\sqrt{l(l+1)}}{b}. 
\end{align}
The function~${\rm Li}_4(z)$ in the second line of (\ref{def:J}) is the polylogarithmic function. 
In the following, we consider a situation in which $e^{-c_l+\beta\mu}\ll 1$ for $l\geq 1$. 
Then the grand potential can be approximated as
\begin{align}
 J(\beta,\mu,{\cal V}_3,{\cal V}_2) &\simeq -\frac{g_{\rm dof}{\cal V}_3}{2\pi^2\beta^4}
 \brc{\pm 2{\rm Li}(\pm e^{\beta\mu})+e^{\beta\mu}Q_1\brkt{\beta\sqrt{\frac{4\pi}{{\cal V}_2}}}}, 
\end{align}
where
\begin{align}
 Q_1(x) &\equiv \sum_{l=1}^\infty x^2l(l+1)(2l+1)K_2\brkt{x\sqrt{l(l+1)}}. 
 \label{def:Q_1}
\end{align}
Here $K_2(z)$ is the modified Bessel function of the second kind. 

From (\ref{def:J}), various thermodynamic quantities are calculated as follows. 
The upper (lower) signs represent the case of bosons (fermions). 
\begin{description}
\item[Radiation energy density] 
\begin{align}
 \rho^{\rm rad} &= \frac{1}{{\cal V}_3{\cal V}_2}\brkt{\partial_\beta-\frac{\mu}{\beta}\partial_\mu}(\beta J) \nonumber\\
 &\simeq \frac{g_{\rm dof}}{2\pi^2\beta^4{\cal V}_2}\brc{\pm 6{\rm Li}_4(\pm e^{\beta\mu})+e^{\beta\mu}\brkt{3Q_1+Q_2}}, 
 \label{expr:rho^rad}
\end{align}
where 
\begin{align}
 Q_2(x) &\equiv -xQ_1'(x) = \sum_{l=1}^\infty x^3l^{3/2}(l+1)^{3/2}(2l+1)K_1\brkt{x\sqrt{l(l+1)}}. 
 \label{def:Q_2}
\end{align}

\item[3D pressure]
\begin{align}
 p^{\rm rad}_3 &= -\frac{1}{{\cal V}_2}\frac{\partial J}{\partial {\cal V}_3}
 \simeq \frac{g_{\rm dof}}{2\pi^2\beta^4{\cal V}_2}\brc{\pm 2{\rm Li}_4(\pm e^{\beta\mu})+e^{\beta\mu}Q_1}. 
 \label{expr:p^rad_3}
\end{align}

\item[2D pressure]
\begin{align}
 p^{\rm rad}_2 &= -\frac{1}{{\cal V}_3}\frac{\partial J}{\partial {\cal V}_2}
 \simeq \frac{g_{\rm dof}e^{\beta\mu}}{4\pi^2\beta^4{\cal V}_2}Q_2. 
 \label{expr:p^rad_2}
\end{align}


\end{description}
The arguments of the functions~$Q_1$ and $Q_2$ are understood as 
$\beta\sqrt{4\pi/{\cal V}_2}=\beta/b$. 

We should note that
\begin{align}
 \rho^{\rm rad} &= 3p^{\rm rad}_3+2p^{\rm rad}_2. \label{rel:rhop}
\end{align}

\begin{figure}[t]
  \begin{center}
    \includegraphics[scale=0.55]{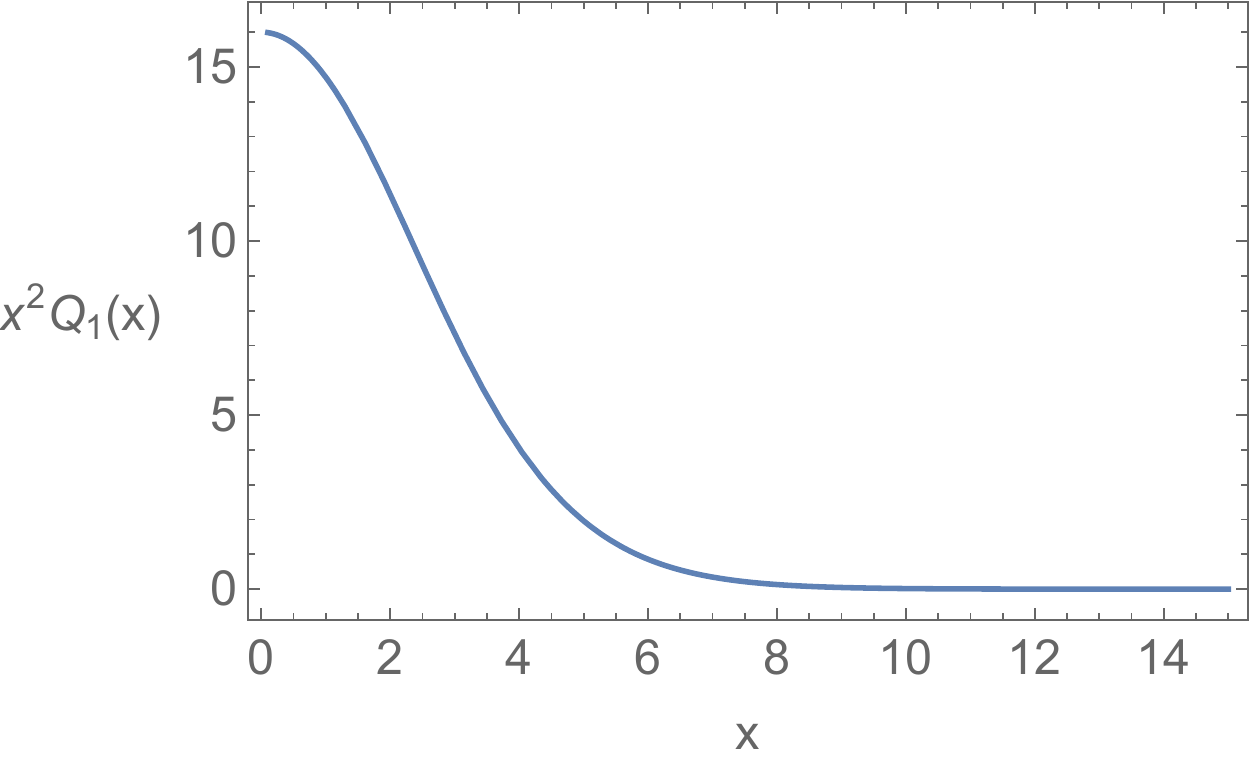} \hspace{10mm}
    \includegraphics[scale=0.55]{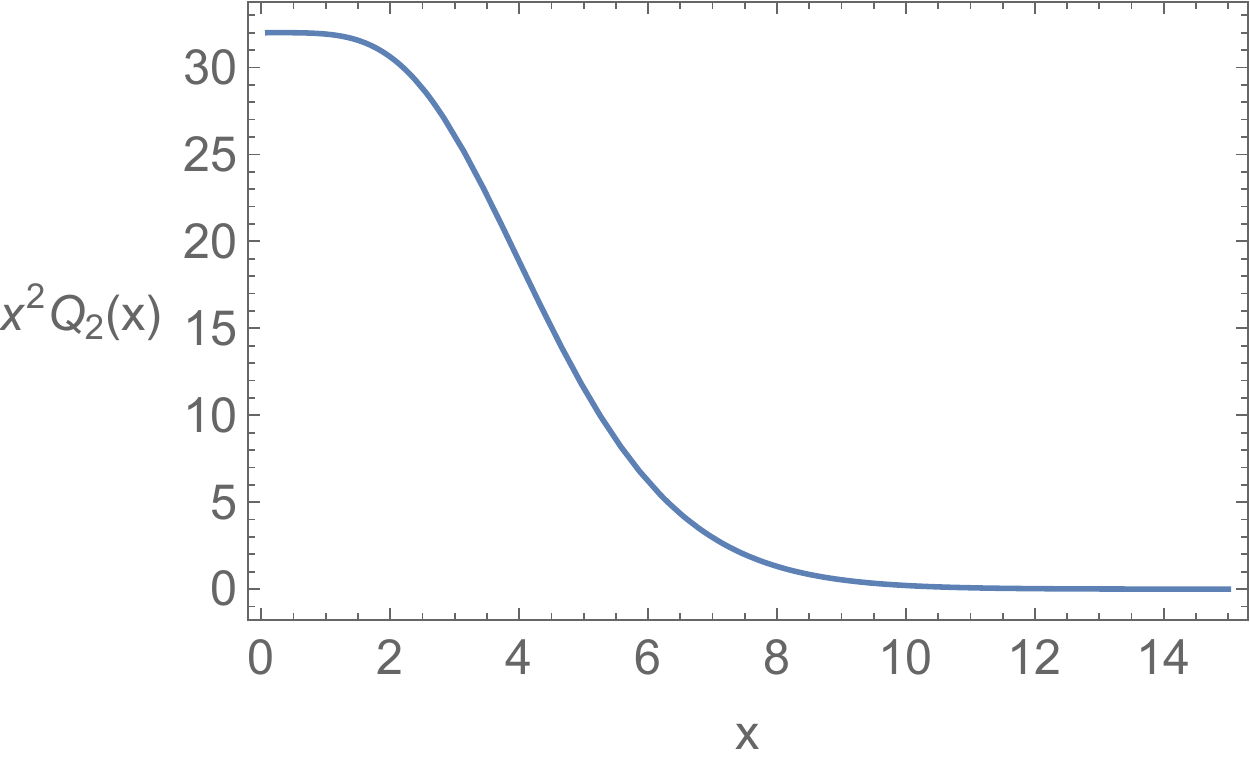} \\
    \includegraphics[scale=0.55]{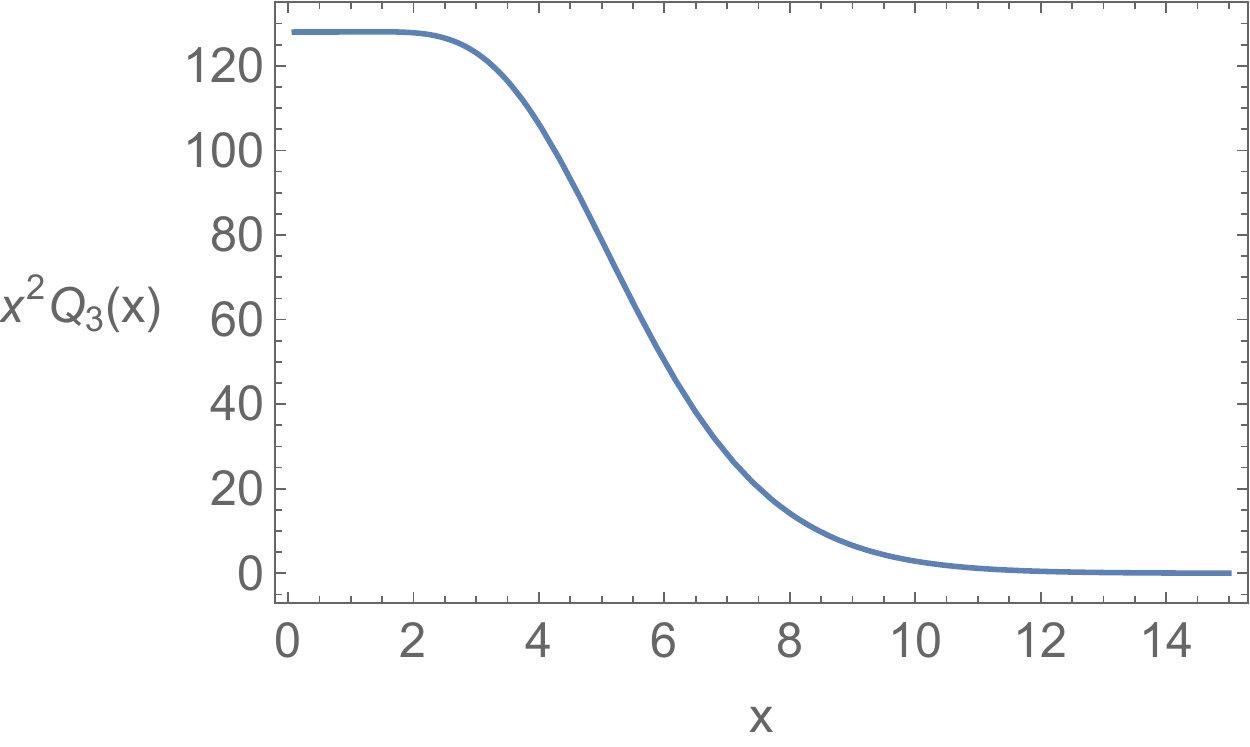} 
  \end{center}
\caption{The profiles of $x^2Q_i(x)$ ($i=1,2,3$). }
    \label{profile:Qs}
\end{figure}

The profiles of the functions~$x^2Q_i(x)$ ($i=1,2,3$) are shown in Fig.~\ref{profile:Qs}. 

\section{Conservation law}
\label{conserv_law}
Including the radiation contribution, the energy-momentum conservation law is 
\begin{align}
 \nabla_M T^M_{\;\;N} &\equiv \partial_M T^M_{\;\;N}+\Gamma^M_{\;\;ML}T^L_{\;\;N}-\Gamma^L_{\;\;MN}T^M_{\;\;L} = 0, 
 \label{T:conserve}
\end{align}
where 
\begin{align}
 T^t_{\;\;t} &= \frac{1}{2}\dot{\sigma}^2+\frac{e^\sigma}{8b^4}+V(\sigma)+\rho^{\rm rad} \equiv \rho^{\rm tot}, \nonumber\\
 T^i_{\;\;j} &= \delta^i_{\;\;j}\brc{-\frac{1}{2}\dot{\sigma}^2+\frac{e^\sigma}{8b^4}+V(\sigma)-p^{\rm rad}_3} 
 \equiv -\delta^i_{\;\;j}p^{\rm tot}_3, \nonumber\\
 T^4_{\;\;4} &= T^5_{\;\;5} = -\frac{1}{2}\dot{\sigma}^2-\frac{e^\sigma}{8b^4}+V(\sigma)-p^{\rm rad}_2
 \equiv -p^{\rm tot}_2. 
 \label{comp:T}
\end{align}
From (\ref{T:conserve}) with $N=t$, we have 
\begin{align}
 \dot{\rho}^{\rm tot}+\frac{3\dot{a}}{a}\brkt{\rho^{\rm tot}+p^{\rm tot}_3}+\frac{2\dot{b}}{b}\brkt{\rho^{\rm tot}+p^{\rm tot}_2} &= 0, 
 \label{rho:conserve}
\end{align}
where the dot denotes the time derivative. 
The other components hold trivially. 
By using the dilaton field equation in (\ref{bg:EOM}), the conservation law~(\ref{rho:conserve}) is reduced to
\begin{align}
 \dot{\rho}^{\rm rad}+\brkt{\frac{3\dot{a}}{a}+\frac{2\dot{b}}{b}}\rho^{\rm rad}
 +\frac{3\dot{a}}{a}p^{\rm rad}_3+\frac{2\dot{b}}{b}p^{\rm rad}_2 &= 0. 
\end{align}
Plugging (\ref{expr:rho^rad}), (\ref{expr:p^rad_3}) and (\ref{expr:p^rad_2}) into this, we obtain 
\begin{align}
 &\frac{\dot{\beta}}{\beta}\left\{
 \pm 24{\rm Li}_4(\pm e^{\beta\mu})+e^{\beta\mu}\brkt{12Q_1+5Q_2+Q_3} \right.\nonumber\\
 &\hspace{5mm}\left. 
 -\beta\mu\brkt{\pm 6{\rm Li}_3(\pm e^{\beta\mu})+e^{\beta\mu}\brkt{3Q_1+Q_2}}\right\} \nonumber\\
 =&\; \frac{3\dot{a}}{a}\brc{\pm 8{\rm Li}_4(\pm e^{\beta\mu})+e^{\beta\mu}\brkt{4Q_1+Q_2}}
 +\frac{\dot{b}}{b}e^{\beta\mu}\brkt{2Q_2+Q_3}, 
 \label{eq_for_beta}
\end{align}
where 
\begin{align}
 Q_3(x) &\equiv 2Q_2(x)-xQ_2'(x) \nonumber\\
 &= \sum_{l=1}^\infty x^4l^2(l+1)^2(2l+1)K_0\brkt{x\sqrt{l(l+1)}}. 
 \label{def:Q_3}
\end{align}
The arguments of $Q_{1,2,3}$ are understood as $\beta/b$.

\section{Derivation of (\ref{Expr:a-x})} 
\label{derivation_a}
From (\ref{rel:dotbt-dota}), we obtain 
\begin{align}
 \int_\beta^{\beta_{j+1}}\frac{d\tl{\beta}}{\tl{\beta}v_\beta^{\rm ap}(\tl{\beta}/b_*)} 
 &= \int_a^{a_{j+1}}\frac{d\tl{a}}{\tl{a}} = \ln\frac{a_{j+1}}{a}, 
 \label{int:bta}
\end{align}
where $a_j\equiv a(\beta_j)$, for $x_j<\beta/b_*\leq x_{j+1}(\leq 10)$. 
The LHS of (\ref{int:bta}) is calculated as
\begin{align}
 \int_{\beta/b_*}^{x_{j+1}}\frac{dx}{xv_\beta^{\rm ap}(x)} 
 &= \sbk{\frac{1}{c_2^{(j)}}\ln\frac{x}{c_1^{(j)}x+c_2^{(j)}}}_{\beta/b_*}^{x_{j+1}} 
 = \frac{1}{c_2^{(j)}}\ln\brkt{{\cal B}_j\frac{c_1^{(j)}\beta+c_2^{(j)}b_*}{\beta}}, 
\end{align}
where
\begin{align}
 {\cal B}_j &\equiv \frac{x_{j+1}}{c_1^{(j)}x_{j+1}+c_2^{(j)}} = \frac{(j+1)\Delta}{c_1^{(j)}(j+1)\Delta+c_2^{(j)}}. 
 \label{def:cB}
\end{align}
Thus, we can express $a$ as a function of $\beta$ as 
\begin{align}
 a(\beta) &= a_{j+1}\brkt{\frac{1}{{\cal B}_j}\frac{\beta}{c_1^{(j)}\beta+c_2^{(j)}b_*}}^{1/c_2^{(j)}}, 
 \label{fct_form:a}
\end{align}
for $x_j<\beta/b_*\leq x_{j+1}(\leq 10)$. 

For $\beta>10b_*$, the corresponding equation to (\ref{int:bta}) is 
\begin{align}
 \ln\frac{a}{a_J} &= \int_{10}^{\beta/b_*}\frac{dx}{x} = \ln\frac{\beta}{10b_*}, 
\end{align}
which leads to 
\begin{align}
 a(\beta) &= \frac{a_J\beta}{10b_*}. 
 \label{fct_form:a:2}
\end{align}

From (\ref{fct_form:a}) and (\ref{fct_form:a:2}), we have
\begin{align}
 A(x) &= \begin{cases} \displaystyle A_{j+1}-\frac{1}{c_2^{(j)}}\ln\brkt{{\cal B}_j\frac{c_1^{(j)}x+c_2^{(j)}}{x}} & (x_j<x\leq x_{j+1}\leq 10) \\
 \displaystyle A_J+\ln\frac{x}{10} & (x>10) \end{cases}, 
 \label{Ax}
\end{align}
where $x\equiv \beta/b_*$ and $A_j\equiv\ln a_j$. 

For the reference time~$t_{\rm ref}$, we define the integer~$k$ so that $x_k <\beta(t_{\rm ref})/b_* \leq x_{k+1}$. 
Then, from (\ref{fct_form:a}), we obtain 
\begin{align}
 a_{\rm ref} &= \frac{a_{k+1}}{K_k(x_{\rm ref})}, 
\end{align}
where $a_{\rm ref}\equiv a(t_{\rm ref})$, $x_{\rm ref}\equiv \beta(t_{\rm ref})/b_*$, and 
\begin{align}
 K_j(x) &\equiv \brkt{{\cal B}_j\frac{c_1^{(j)}x+c_2^{(j)}}{x}}^{1/c_2^{(j)}} 
 = \brkt{\frac{x_{j+1}}{c_1^{(j)}x_{j+1}+c_2^{(j)}}\cdot \frac{c_1^{(j)}x+c_2^{(j)}}{x}}^{1/c_2^{(j)}}. 
 \label{def:K_j}
\end{align}
Therefore, using (\ref{fct_form:a}) repeatedly, we have 
\begin{align}
 a(x) &= \frac{a_{k+1}}{K_k(x)} = \frac{a_{\rm ref}K_k(x_{\rm ref})}{K_k(x)}, 
 \label{expr:a:k}
\end{align}
for $x_{\rm ref}\leq x\leq x_{k+1}$, 
\begin{align}
 a(x) &= \frac{a_{k+2}}{K_{k+1}(x)} = \frac{a_{k+1}K_{k+1}(x_{k+1})}{K_{k+1}(x)} 
 = \frac{a_{\rm ref}K_k(x_{\rm ref})K_{k+1}(x_{k+1})}{K_{k+1}(x)}, 
\end{align}
for $x_{k+1}\leq x\leq x_{k+2}$, 
\begin{align}
 a(x) &= \frac{a_{j+1}}{K_j(x)} = a_jK_j(x_j)\frac{1}{K_j(x)} = a_{j-1}K_{j-1}(x_{j-1})K_j(x_j)\frac{1}{K_j(x)} \nonumber\\
 &= \cdots = a_{k+1}K_{k+1}(x_{k+1})\cdots K_j(x_j)\frac{1}{K_j(x)} \nonumber\\
 &= a_{\rm ref}K_k(x_{\rm ref})K_{k+1}(x_{k+1})\cdots K_j(x_j)\frac{1}{K_j(x)}, 
 \label{expr:a:1}
\end{align}
for $x_j<x\leq x_{j+1}\leq x_J=10$, and (\ref{fct_form:a:2}) is expressed as 
\begin{align} 
 a(x) &= a_{\rm ref}\frac{x}{10}K_k(x_{\rm ref})K_{k+1}(x_{k+1})\cdots K_{J-1}(x_{J-1}), 
 \label{expr:a:2}
\end{align}
for $x>10$.


\end{document}